\documentclass[12pt,a4paper]{article}
\pdfoutput=1
\usepackage{jheppub}
 \usepackage[utf8]{inputenc}
\usepackage{graphicx}
\usepackage{lipsum}
\usepackage[all]{hypcap}
\usepackage{multirow}
\usepackage{rotating}
\usepackage{float}
\usepackage[section]{placeins}
\usepackage{bbold} % For the identity symbol (\mathbb{1})
\usepackage{slashed}

\newcommand{\vv}{\vspace{2mm}}

\begin{document}

\renewcommand{\arraystretch}{1.2}

\title{Flavored Dark Sectors running to low energy}

\affiliation[]{Instituto de F\'{i}sica, Universidade de S\~{a}o Paulo, \\C.P. 66.318, 05315-970 S\~{a}o Paulo, Brazil}

\author[]{Mart\'{i}n Arteaga,}
\author[]{Enrico Bertuzzo,}
\author[]{Cristian Caniu Barros,}
\author[]{Zahra Tabrizi}

\emailAdd{martin77@if.usp.br}
\emailAdd{bertuzzo@if.usp.br}
\emailAdd{caniu@if.usp.br}
\emailAdd{ztabrizi@if.usp.br}

\abstract{
We consider the effective field theory generated by a heavy mediator that connects Standard Model particles to a Dark Sector, considering explicitly the flavor structure of the operators. In particular, we study the model independent running and mixing between operators, as well as their matching at the electroweak scale. In addition to the explicit expression of the Renormalization Group Equations, we show the numerical solutions as well as some approximate analytical expressions that help understanding these solutions. At low energy, our results are particularly important in the case of light dark sectors communicating to the $b$ quark, and can be immediately applied to flavored Dark Matter.
}

\maketitle

\section{Introduction}

Over the last few years, dark sectors have started to play an increasingly important role in Beyond the Standard Model (BSM) physics. The reason is twofold: on one hand, the null results from the LHC and other experiments are pushing `traditional' BSM theories to more and more tuned regions of their parameter space, motivating the search for new unconventional signatures~\cite{Strassler:2006im,Strassler:2006ri, Beauchesne:2017yhh}. On the other hand, dark sectors are implicitly present in many Dark Matter (DM) models (especially in connection with light mediators, see for instance~\cite{Battaglieri:2017aum}), and are starting to appear in more recent solutions of the Hierarchy Problem like Twin Higgs~\cite{Chacko:2005pe}, Folded SUSY~\cite{Burdman:2006tz} and the relaxation of the electroweak (EW) scale~\cite{Graham:2015cka}. In addition, many dark sectors predict the existence of Long-Lived Particles (LLPs), for which a new extensive experimental program is being developed (see for instance~\cite{Curtin:2018mvb}). \vv

Of course, given the plethora of possible dark sectors with different symmetries and particles, a general analysis is impossible, and some broad assumption must be made. What we consider in this paper is a generic dark sector, communicating with the SM via some unspecified mediator $X$ whose interactions are allowed to be flavor off-diagonal. If the mediator is somewhat heavier than the typical energy scale of low energy experiments (say a few GeV), then the mediator can be integrated out, generating effective contact interactions of the type
\begin{align}\label{eq:product_currents}
    \mathcal{L}_{EFT} \supset c J_{\cal D} J_{SM}\, ,
\end{align}
where $J_{\cal D}$ and $J_{SM}$ are currents involving, respectively, dark and SM fields only, $c$ is an appropriate coefficient ensuring the right dimensions, and we do not write possible Lorentz indices. This generic framework resembles typical Hidden Valley constructions~\cite{Strassler:2006im,Strassler:2008fv}, and can be obtained with a variety of mediators~\cite{Beauchesne:2017yhh}. Also, the nature of the dark current depends crucially on the details of the dark sector, as it can be composed by elementary or composite fields. Since we will remain agnostic about the details of the dark sector, we dub the Effective Field Theory (EFT) defined in Eq.~\eqref{eq:product_currents} by `dark sector EFT'. Specific cases have already been analyzed in the literature, as we are going to see.\vv

In order to compare the Wilson coefficients in Eq.~\eqref{eq:product_currents} with experimental data, care must be taken with the fact that renormalization group effects can (and generically will) be important. The running and mixing of operators written in the form of Eq.~\eqref{eq:product_currents} has been studied in detail in the case in which $J_{SM}$ is diagonal in flavor space and $J_{\cal D}$ is a Dark Matter current~\cite{Hill:2011be,Frandsen:2012db,Vecchi:2013iza,Crivellin:2014qxa,DEramo:2014nmf,DEramo:2016gos,Brod:2017bsw,Brod:2018ust,Belyaev:2018pqr,Beauchesne:2018myj,Chao:2016lqd}, and has been applied to the case of DM with mass in the MeV range in Ref.~\cite{Bertuzzo:2017lwt}. Some code has also been publicly released~\cite{runDM,Bishara:2017nnn}, computing numerically the solution to the Renormalization Group Equations (RGE's). As a matter of fact, as long as we do not consider dark particles in the loops, the results of Refs.~\cite{DEramo:2016gos,Bishara:2017nnn} can be applied to any dark sector coupling to the SM via some flavor-blind heavy mediator, and not only to the case of Dark Matter currents. \vv

The purpose of this paper is to extend the analysis of the running and mixing of the dark sector EFT operators \textit{supposing the mediator to be heavy and flavorful}. Such a mediator has already been considered in the framework of flavored Dark Matter (see for instance~\cite{1104.5239,1105.1781,1107.0623,1109.3516,1303.1087,1308.0584,1308.0612,1309.4462,1404.1918,1405.6709,1612.01643,Blanke:2017tnb,1711.10493,1712.00037}), but the running of the EFT has never been considered before. We keep our analysis as general as possible, without committing to a specific dark sector or requiring Dark Matter to be present. We restrict our analysis to the mixing of effective operators written as in Eq.~\eqref{eq:product_currents}. Of course, this is still not the complete renormalization of the dark sector EFT. Once the dimension of the operators to be included in the Lagrangian is fixed (\textit{i.e.} the nature of the dark current is specified), we also need to include operators constructed out of SM fields only. In addition, loops of dark particles will contribute to the renormalization of the complete EFT, and the effect may be important. This has been recently studied in the context of Dirac Fermion DM EFT in Ref.~\cite{Bishara:2018vix}, and in the case of flavorful mediators could, for instance, generate contributions to meson mixing operators. The information is of course fundamental for the comparison between data and theory but depends strongly on the nature of the particles running in the loops. As such, we decided to focus here only on the \textit{model independent} information that can be extracted from the running and mixing of the dark sector EFT operators. We will study elsewhere the effects of dark particles in the loops.\vv
 
The paper is organized as follows: in Section~\ref{sec:DSEFT} we build the dark sector Effective Field Theory, describing the operators considered above and below the EW scale. In Section~\ref{sec:RGEs} we discuss the RGE's and show the matching needed in evolving the Wilson coefficients from high to low energy. Finally, Section~\ref{sec:results} is devoted to the numerical results and to possible phenomenological applications. We give our conclusions in Section~\ref{sec:summary}. 

\begin{table}
\centering
\begin{tabular}{|c|c|c|c|c|c|c|c}
\hline
	 & $q_{L}^i$ & $u_R^i$ & $d_R^i$ & $l_L^i$ & $e_R^i$ & $H$ \\
\hline
\hline
$SU(3)_c$& \textbf{3} & \textbf{3} & \textbf{3} & \textbf{1} & \textbf{1} & \textbf{1}\\
$SU(2)_L$& \textbf{2} & \textbf{1} & \textbf{1} & \textbf{2} & \textbf{1} & \textbf{2}\\
$U(1)_Y$ & +1/6 & +2/3 & -1/3 & -1/2 & -1 & +1/2\\
\hline
\end{tabular}
\caption{\label{field_rep} Charges and gauge representations of the SM fields above the EW scale. The index $i$ is a family index.}
\end{table}

%%%%%%%%%%%%%%%%%%%%%%%%%%%
\section{Dark Sector Effective Field Theory}\label{sec:DSEFT}
%%%%%%%%%%%%%%%%%%%%%%%%%%%

%

We start in this Section with the definition of our framework. The effective Lagrangian contains the SM Lagrangian, the kinetic term for the dark fields and the interactions between the dark sector and the SM particles. As explained in the Introduction, we will work with interactions which are products of a dark and of a SM current,
\begin{align}\label{eq:EFT_Lagr}
    \mathcal{L}_{int} \supset \frac{(C_a)_{ij}}{\Lambda^n} J_{\cal D}^a J_{SM}^{a, (ij)} + h.c.
\end{align}
The index $a$ runs over all possible currents, while $(i,j)$ are SM family indices. We do not write explicitly possible Lorentz indices. The dark currents can have a variety of forms, depending on the nature of the dark sector. For instance, they can be constructed out of fundamental fermions (like $J_{\cal D}^\mu = \overline{\chi}\gamma^\mu \chi$ or $J_{\cal D} = \overline{\chi}\chi$), out of fundamental scalars (like $J_{\cal D}^\mu = \phi^\dag \overleftrightarrow{\partial}^\mu \phi$ or $J_{\cal D} = \phi^\dag \phi$), or they can be constructed out of composite objects (pions or baryons) if the dark sector is strongly interacting at low energies~\cite{Hochberg:2014dra,Okawa:2016wrr,Beauchesne:2018myj}. Depending on the dimension of the currents, the appropriate $n$ must be chosen. Since we do not want our conclusions to depend on the nature of the dark sector, in this work we will leave unspecified the nature of the dark current, making only the broad assumption that it is a complete gauge singlet.  Our conventions for the SM fields are presented in Table~\ref{field_rep} and in Appendix~\ref{app:details}.\vv

\begin{table}[t!b]
    \centering
    \begin{tabular}{cccc}
        dimensions & currents & Wilson coefficient & flavor transformation \\
        \hline \hline
        \multirow{3}{*}{$d=2$}  & $H^\dag H$ & $C_{H\;portal}$ & singlet \\
                                & $B_{\mu\nu}$ &  $C_{Y\;portal}$ & singlet\\
                                & $\tilde{B}_{\mu\nu}$ & $\tilde{C}_{Y\;portal}$ & singlet\\
        \hline
        $d=5/2$                & $\overline{\ell}_L \tilde{H}$ & $C_{N\; portal}$ & $\mathbf{3}$ of $SU(3)_{\ell_L}$ \\
        \hline
        \multirow{6}{*}{$d=3$} & $\overline{q}_L \gamma^\mu q_L$ & $C_{q_L}$ & $(\mathbf{3}, \overline{\mathbf{3}})$ of $SU(3)_{q_L} \times SU(3)_{q_L}$ \\
                               & $\overline{u}_R \gamma^\mu u_R$ & $C_{u_R}$ & $(\mathbf{3}, \overline{\mathbf{3}})$ of $SU(3)_{u_R} \times SU(3)_{u_R}$ \\
                               & $\overline{d}_R \gamma^\mu d_R$ & $C_{d_R}$ & $(\mathbf{3}, \overline{\mathbf{3}})$ of $SU(3)_{d_R} \times SU(3)_{d_R}$ \\
                               & $\overline{\ell}_L \gamma^\mu \ell_L$ & $C_{\ell_L}$ & $(\mathbf{3}, \overline{\mathbf{3}})$ of $SU(3)_{\ell_L} \times SU(3)_{\ell_L}$ \\
                               & $\overline{e}_R \gamma^\mu e_R$ & $C_{e_R}$ & $(\mathbf{3}, \overline{\mathbf{3}})$ of $SU(3)_{e_R} \times SU(3)_{e_R}$ \\
                               & $i H^\dag \overleftrightarrow{D}_\mu H$ & $C_H$ & singlet \\
                               & $\partial_\nu B^{\nu\mu}$ & $C_B$ & singlet\\
        \hline
        % \multirow{3}{*}{$d=4$} & $\overline{q}_L H d_R$ & $C_{Y_d}$ & $(\mathbf{3}, \overline{\mathbf{3}})$ of $SU(3)_{q_L} \times SU(3)_{d_R}$ \\
        %                         & $\overline{q}_L \tilde{H} u_R$ & $C_{Y_u}$ & $(\mathbf{3}, \overline{\mathbf{3}})$ of $SU(3)_{q_L} \times SU(3)_{u_R}$ \\
        %                         & $\overline{\ell}_L H e_R$ & $C_{Y_e}$ & $(\mathbf{3}, \overline{\mathbf{3}})$ of $SU(3)_{\ell_L} \times SU(3)_{e_R}$ \\
        % \hline
    \end{tabular}
    \caption{\label{tab:SM_currents_aboveEW} List of $d<4$ currents constructed out of SM fields to be used in Eq.~\eqref{eq:EFT_Lagr} only above the EW scale. We have suppressed all flavor indices. In our analysis we will focus on the running and mixing of $d=3$ currents.}
\end{table}
Let us first consider the EFT \textit{above} the Electroweak scale. In this case, we demand the SM currents $J_{SM}^a$ to be complete gauge singlets under $SU(3)_c \times SU(2)_L \times U(1)_Y$, with particle contents and charges given in Table \ref{field_rep}. The assumption can of course be relaxed, see for instance Ref.~\cite{Brod:2017bsw}. The SM currents can be classified according to their dimensions, while the coefficients $(C_a)_{ij}$ in Eq.~\eqref{eq:EFT_Lagr} can be classified based on their transformation properties under the SM flavor group $SU(3)_{q_L} \times SU(3)_{u_R} \times SU(3)_{d_R} \times SU(3)_{\ell_L} \times SU(3)_{e_R}$ (explicitly broken by the Yukawa couplings). This is shown in Table~\ref{tab:SM_currents_aboveEW}. At the level of $d=2$ and $d=5/2$ we have the scalar Higgs, hypercharge and neutrino portals, which are the only currents that allow for renormalizable portals between the SM and the dark sector. These currents have been extensively used in the context of sub-GeV Dark Matter (see for instance~\cite{Burgess:2000yq,ArkaniHamed:2008qn,Falkowski:2009yz}) and, more recently, for the generation of neutrino masses~\cite{Bertuzzo:2018ftf}. From the point of view of running, none of these currents mix with the others. As long as the nature of the dark sector is not specified, the only relevant effects would be the thresholds corrections coming from wave function renormalization, that can be easily computed using Appendix~\ref{app:computation_RGE}.
Since in this paper we will focus on the case of heavy mediators (\textit{i.e.} on the case of non-renormalizable interactions between the SM and the dark sector), we will not consider these currents in the rest of the paper. Moving on, non-trivial structures with both lepton and quark flavors appear at the level of $d=3$, and these are the currents on which we will focus from now on. Of course, more currents with non-trivial flavor structure appear with $d\geq 4$, but since their effects are suppressed by higher powers of $\Lambda$, we will not consider them in the remainder of the paper. 

It is interesting to count the number of parameters in the dark sector EFT, focusing on the $d=3$ currents. For the fermion bilinears, the Wilson coefficients shown in Table~\ref{tab:SM_currents_aboveEW} are $3\times 3$ matrices in flavor space. Moreover, it should be noted that the current $\partial_\nu B^{\nu\mu}$ is redundant, since it can be eliminated using the Equation of Motion (EoM) of the hypercharge field (see Appendix~\ref{app:details}). This leaves us with a total of 46 independent currents to be probed above the EW scale. \vv
\begin{table}[tb]
    \centering
    \begin{tabular}{cccc}
        dimensions & currents & Wilson coefficient & flavor transformation \\
        \hline \hline
        $d=3/2$ & $\nu_L$ & $C_{\nu \; dark}$ & $\mathbf{3}$ of $SU(3)_{\nu_L}$ \\
        \hline
        \multirow{2}{*}{$d=2$}  & $F_{\mu\nu}$ &  $C_{A\;portal}$ & singlet\\
                                & $\tilde{F}_{\mu\nu}$ & $\tilde{C}_{A\;portal}$ & singlet\\
        \hline
        \multirow{6}{*}{$d=3$} & $\overline{u} \gamma^\mu u$ & $C_{V_u}$ & $(\mathbf{2}, \overline{\mathbf{2}})$ of $SU(3)_{u} \times SU(3)_{u}$ \\
                               & $\overline{d} \gamma^\mu d$ & $C_{V_d}$ & $(\mathbf{3}, \overline{\mathbf{3}})$ of $SU(3)_{d} \times SU(3)_{d}$ \\
                               & $\overline{e} \gamma^\mu e$ & $C_{V_e}$ & $(\mathbf{3}, \overline{\mathbf{3}})$ of $SU(3)_{e} \times SU(3)_{e}$ \\
                               & $\overline{\nu_L} \gamma^\mu \nu_L$ & $C_{V_{L\nu}}$ & $(\mathbf{3}, \overline{\mathbf{3}})$ of $SU(3)_{\nu_L} \times SU(3)_{\nu_L}$ \\
                               & $\overline{u} \gamma^\mu\gamma_5 u$ & $C_{A_u}$ & $(\mathbf{2}, \overline{\mathbf{2}})$ of $SU(3)_{u} \times SU(3)_{u}$ \\
                               & $\overline{d} \gamma^\mu\gamma_5 d$ & $C_{A_d}$ & $(\mathbf{3}, \overline{\mathbf{3}})$ of $SU(3)_{d} \times SU(3)_{d}$ \\
                               & $\overline{e} \gamma^\mu \gamma_5 e$ & $C_{A_e}$ & $(\mathbf{3}, \overline{\mathbf{3}})$ of $SU(3)_{e} \times SU(3)_{e}$ \\
                               & $\partial_\nu F^{\nu\mu}$ & $C_\gamma$ & singlet\\
        \hline
    \end{tabular}
    \caption{\label{tab:SM_currents_belowEW} List of $d<4$ currents constructed out of SM fields to be used in Eq.~\eqref{eq:EFT_Lagr} only below the EW scale. $F_{\mu\nu}$ denotes the photon field strength, while $u$, $d$ and $e$ are Dirac fermions. We have suppressed all flavor indices.}
\end{table}
\\

Moving to the EFT \textit{below} the EW scale, we use Dirac fermions to construct currents that are gauge singlets under $SU(3)_c \times U(1)_{em}$. The possible currents up to dimension $3$ are shown in Table~\ref{tab:SM_currents_belowEW}. The flavor symmetry is now $SU(2)_u \times SU(3)_d \times SU(3)_e \times SU(3)_{\nu_L}$, explicitly broken by fermion masses, leaving a total of 53 independent currents to be probed below the EW scale. Notice that we do not introduce right-handed neutrinos in the low energy EFT, and we leave unspecified the mechanism behind neutrino masses. As we did above the EW scale, we will focus in the following on the running and mixing of the $d=3$ currents, again ignoring all possible renormalizable portals.\vv

Our set up will be the following. We will assume the operators in Eq.~\eqref{eq:EFT_Lagr} to be generated at some scale $\Lambda$, to be roughly identified with the mass of some flavorful mediator. We present in Appendix~\ref{app:models} some specific examples using models present in the literature, see Eq.~\eqref{eq:explicit_EFT}. If $\Lambda > \Lambda_{EW}\simeq m_Z$, we will use the $d=3$ SM currents presented in Table~\ref{tab:SM_currents_aboveEW}, while if $\Lambda < \Lambda_{EW}$ we will use the $d=3$ currents from Table~\ref{tab:SM_currents_belowEW}. In both cases, we will leave the flavor structure of the Wilson coefficients $(C_a)_{ij}$ completely generic. In the next section we will compute the running and mixing of such currents from the scale $\Lambda$ at which the operators are generated down to $E \ll \Lambda_{EW}$.

%%%%%%%%%%%%%%%%%%%%%%%%%%%%%%
\section{Renormalization group equations for the dark sector EFT}\label{sec:RGEs}
%%%%%%%%%%%%%%%%%%%%%%%%%%%%%%
\begin{figure}[t!b]
    \centering
    \includegraphics[width = .3 \textwidth]{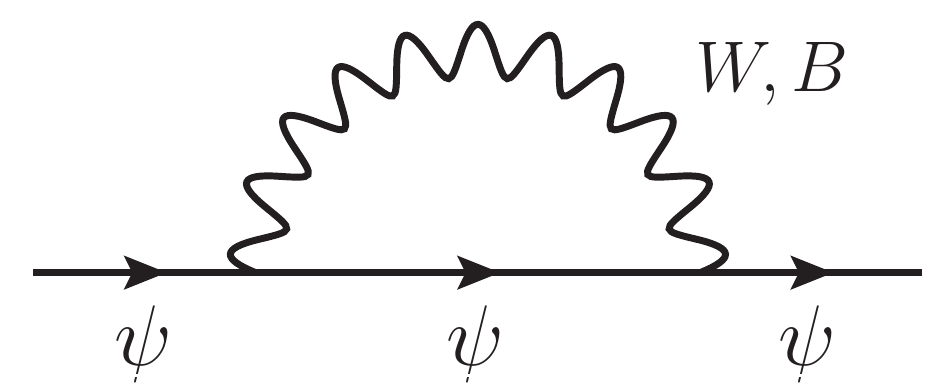} 
    \includegraphics[width = .3 \textwidth]{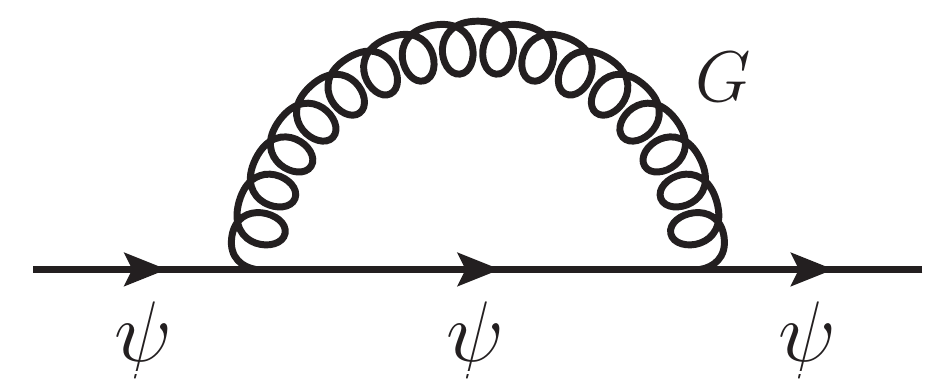}
    \includegraphics[width = .3 \textwidth]{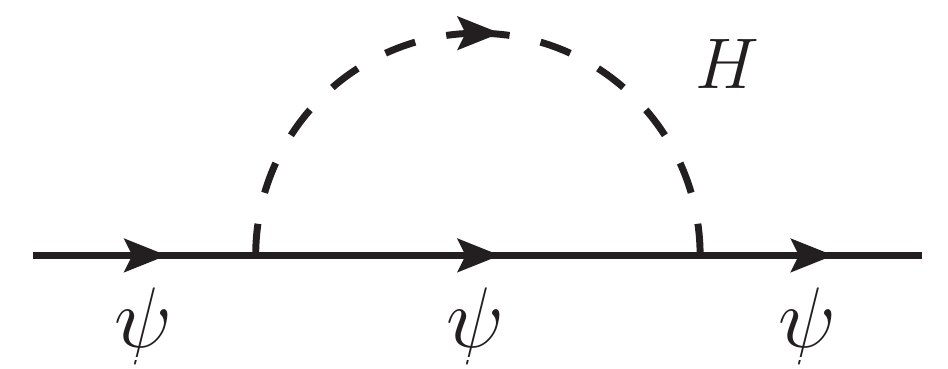}\\ 
    \includegraphics[width = .3 \textwidth]{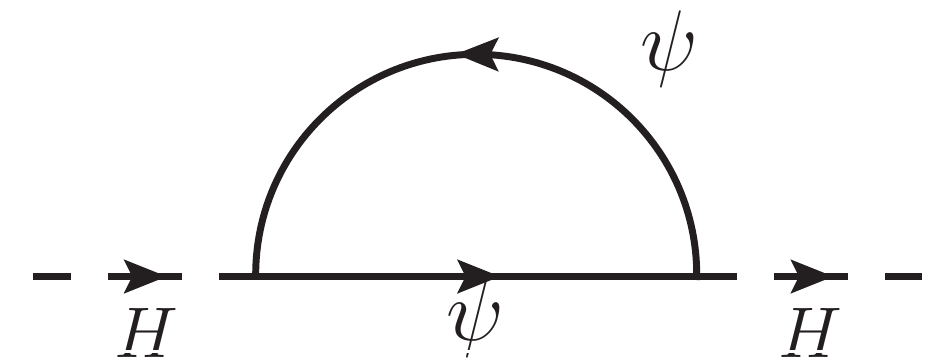}
    \includegraphics[width = .3 \textwidth]{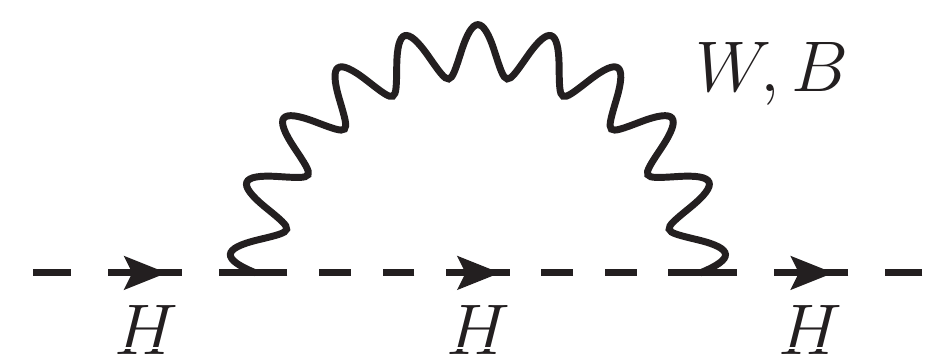}    
    \includegraphics[width = .22 \textwidth]{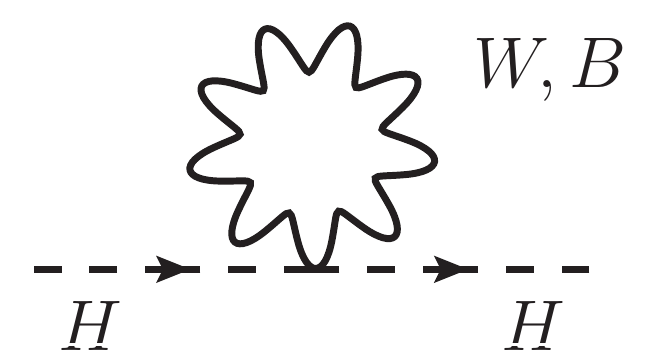}
    \caption{\label{fig:wave_function} Feynman diagrams contributing to the wave function renormalization of fermions $\psi$ and of the Higgs doublet $H$.}
\end{figure}
We start considering the EFT above the EW scale. To be explicit, the Lagrangian we consider is
\begin{align}\begin{aligned}\label{eq:EFT_above}
    \mathcal{L} &= \frac{1}{\Lambda^n} J_{\mathcal{D}\mu} \bigg[ \overline{q}_L \gamma^\mu C_{q_L} q_L + \overline{u}_R \gamma^\mu C_{u_R} u_R + \overline{d}_R \gamma^\mu C_{d_R} d_R  \\ 
                & \qquad {} + \overline{\ell}_L \gamma^\mu C_{\ell_L} \ell_L + \overline{e}_R \gamma^\mu C_{e_R} e_R + C_H i H^\dag \overleftrightarrow{D}_\mu H\bigg]\, ,
\end{aligned}\end{align}    
where all the Wilson coefficients except $C_H$ are matrices in flavor space. To be conservative, we suppose that the same dark current is coupled to all the SM terms, but it is easy to extend the analysis to more general cases. Notice that we do not include the current $\partial_\nu B^{\nu\mu}$ since it is redundant (See Appendix \ref{app:details}).%(it can be eliminated using the equations of motion for the vector $B$, see Eq.~\eqref{eq:EOM_1}). 
\begin{figure}[tb]
    \centering
    \includegraphics[width = .18 \textwidth]{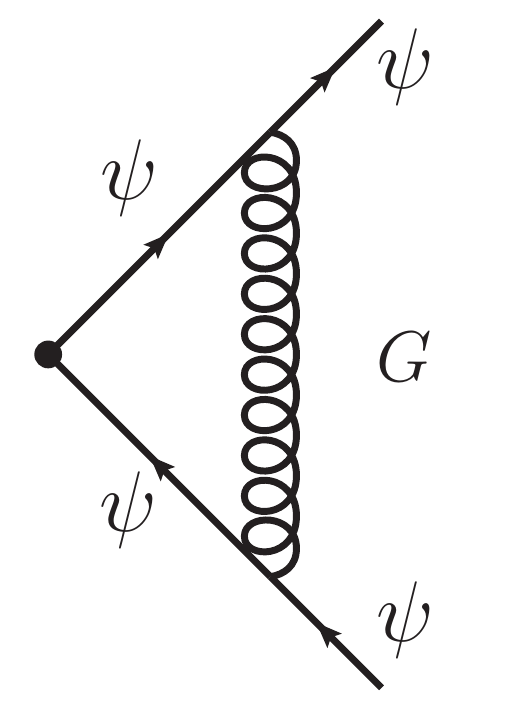} 
    \includegraphics[width = .18 \textwidth]{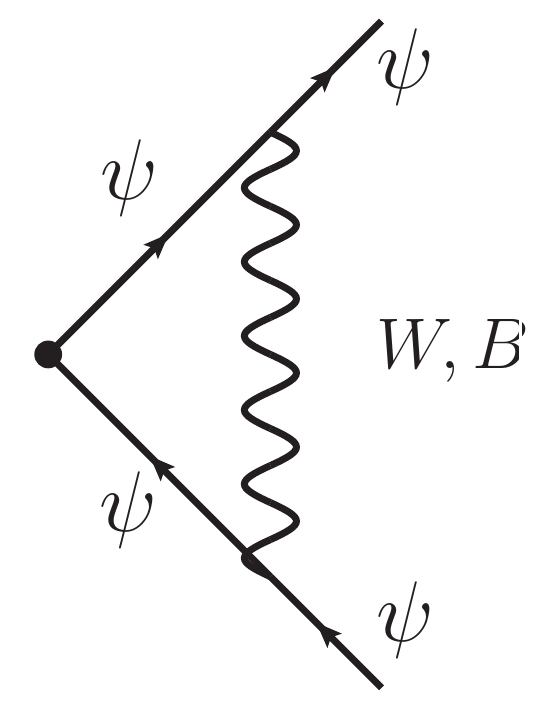}
    \includegraphics[width = .18 \textwidth]{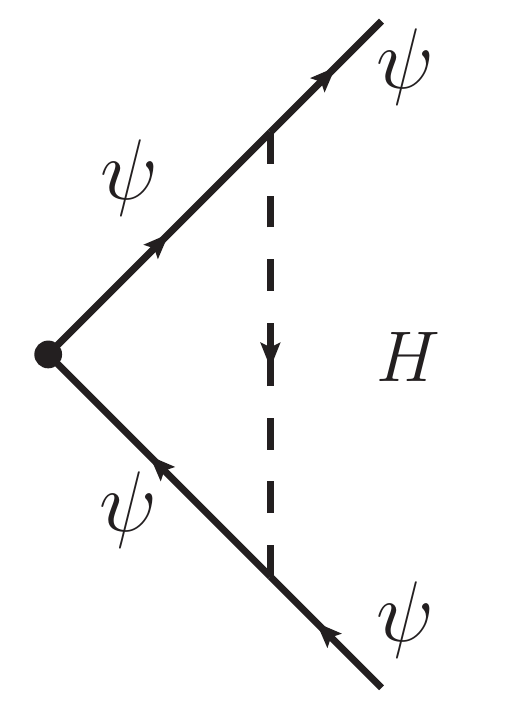} 
    \includegraphics[width = .18 \textwidth]{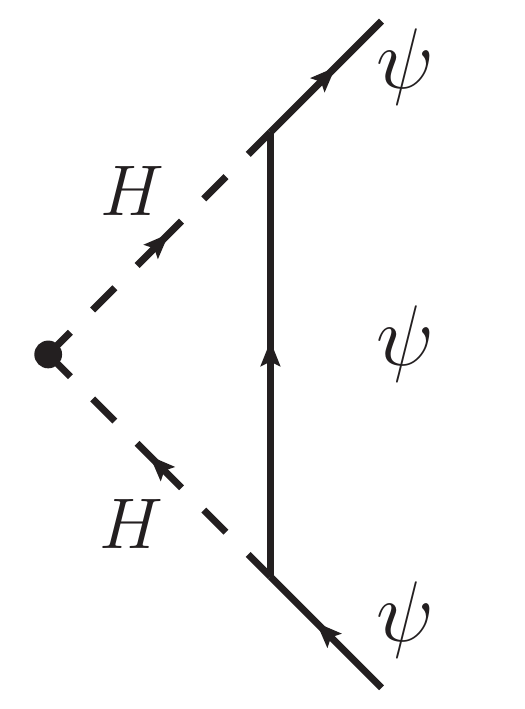}\\
    \includegraphics[width = .18 \textwidth]{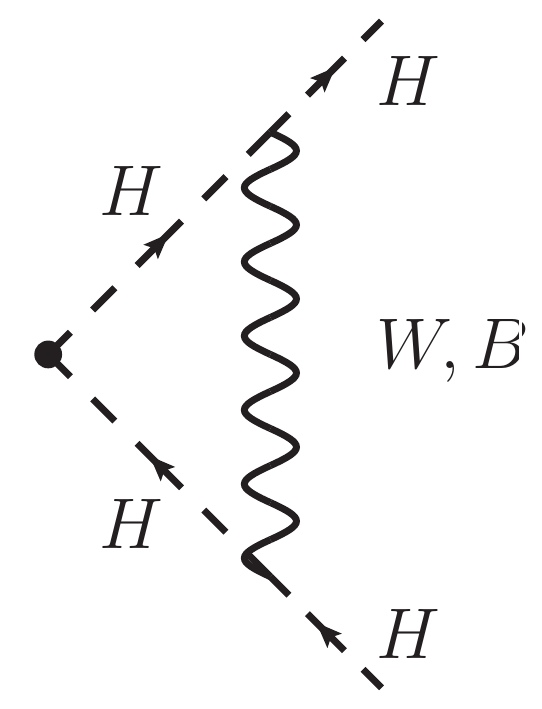}    
    \includegraphics[width = .18 \textwidth]{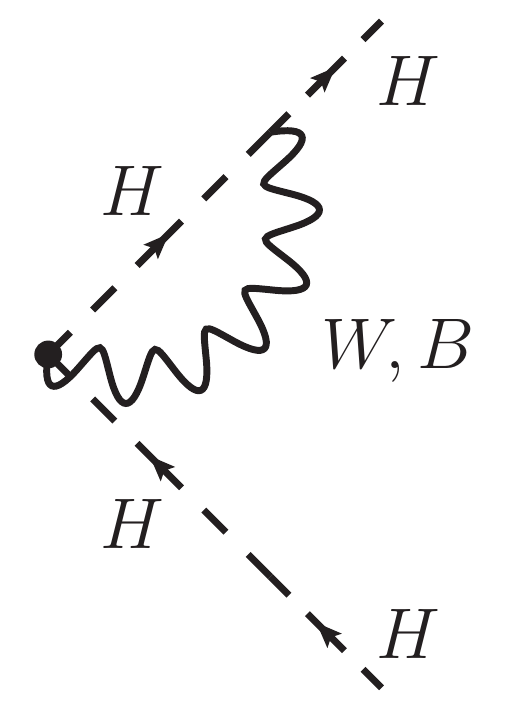}    
    \includegraphics[width = .18 \textwidth]{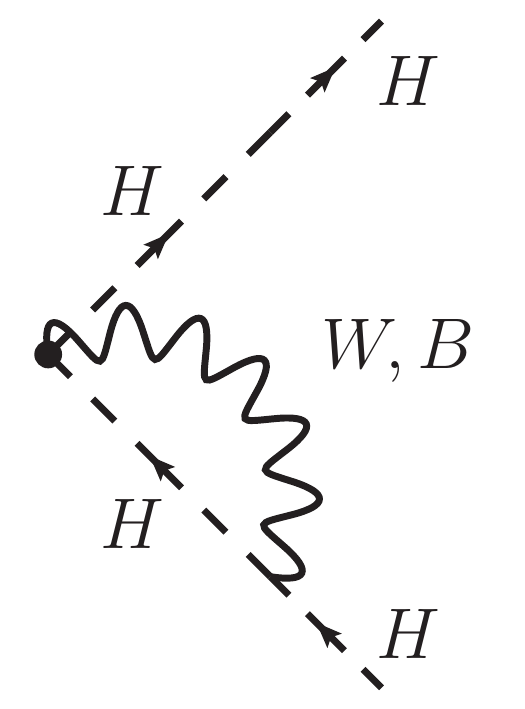}    
    \includegraphics[width = .18 \textwidth]{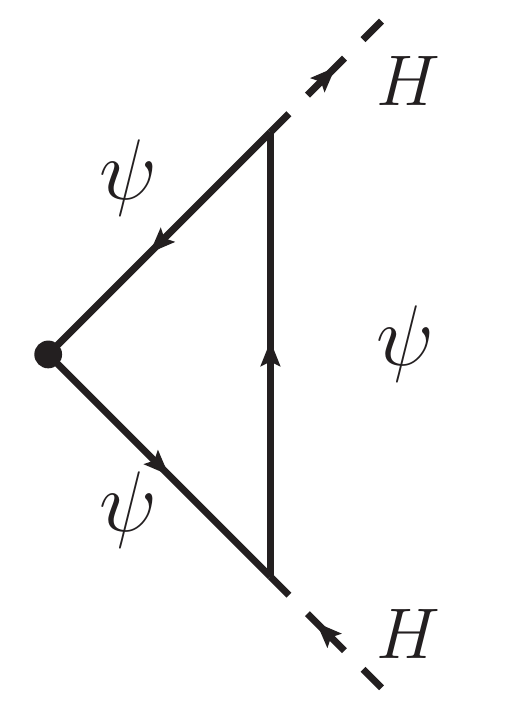}
    \caption{\label{fig:ver_corr} Feynman diagrams contributing to the current corrections for pure fermions $\psi$ and Higgs currents.}
\end{figure}

\begin{figure}[tb]
    \centering
    \includegraphics[width = .25 \textwidth]{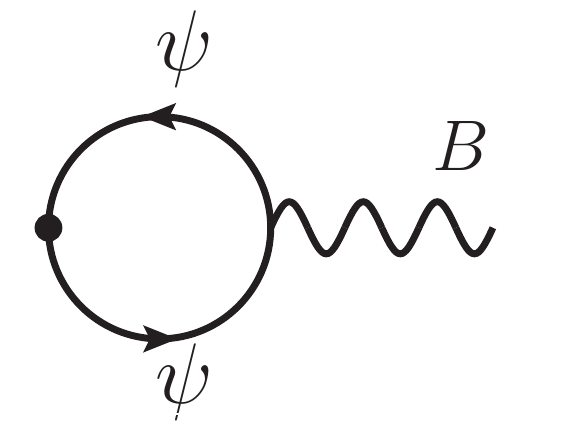} 
    \includegraphics[width = .25 \textwidth]{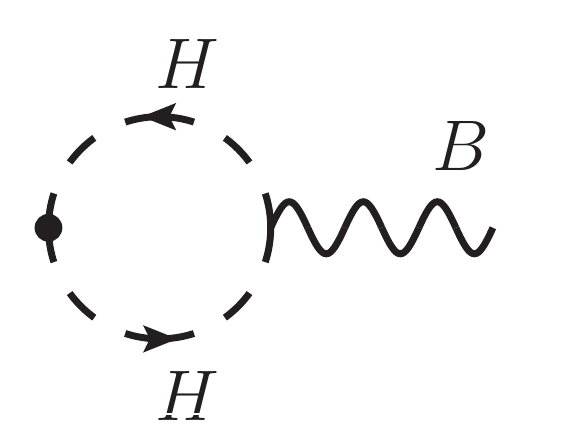}
    \caption{\label{fig:redop_corr} Feynman diagrams contributing to the redundant current $\partial_\nu B^{\nu\mu}$.}
\end{figure}
Since the dark sector is a gauge singlet of the SM symmetry, only the SM fields and interactions are involved in the computation of the RGE's. Here we take into account the wave function renormalization of the fermion and Higgs fields (see Fig.~\ref{fig:wave_function}), as well as the `pure' current corrections shown in Fig.~\ref{fig:ver_corr}. %The top left and middle diagrams in Fig.~\ref{fig:wave_function} give the gauge interaction contributions to the wave function renormalization,  the top right diagram gives correction due to the Yukawa interactions and the bottom diagrams are external leg corrections to the SM Higgs. 
%{\textcolor{red}{The corrections by gauge diagrams will explicitly cancel with the gauge contributions coming from the vertex corrections induced by blah and blah in Fig. \ref{fig:ver_corr}}}.
The top diagrams in Fig.~\ref{fig:ver_corr} are corrections to the fermion currents due to gauge and Yukawa interactions, while the bottom diagrams represent the loop contributions to the Higgs current. Divergencies induced by gauge interactions in the self-energy contributions and in the vertex corrections explicitly cancel one against the other in the final result, leaving only corrections proportional to the Yukawa matrices squared. However, additional care must be taken since, as shown in Fig~\ref{fig:redop_corr}, radiative corrections generate the redundant current $\partial_\nu B^{\nu\mu}$. This contribution must be redefined away again using the equation of motion after the theory has been renormalized. For each Wilson coefficient, this produces an extra correction proportional to the hypercharge $y_\psi$. More details are presented in Appendix~\ref{app:computation_RGE}. The complete computation gives
 
\begin{eqnarray}\label{eq:RGE_above}
(4\pi)^2\frac{dC_{q_L}}{d\log\mu} &=&\frac{1}{2} \left[C_{q_L} Y^2_q + Y^2_q C_{q_L}\right]
                     - \left[Y_u C_{u_R} Y_u^\dagger + Y_d C_{d_R} Y_d^\dagger - Y^2_qC_{H}\right]+y_{q_L} T\mathbb{1}  , \, \nonumber\\
(4\pi)^2\frac{dC_{u_R}}{d\log\mu}  &=& \left[C_{u_R} Y_u^\dagger Y_u + Y_u^\dagger Y_u C_{u_R}\right]
                             -2\left[Y_u^\dagger C_{q_L} Y_u - Y_u^\dagger Y_u C_{H}\right] + y_{u_R} T\mathbb{1}\, , \nonumber\\
(4\pi)^2\frac{dC_{d_R}}{d\log\mu}  &=& \left[C_{d_R} Y_d^\dagger Y_d + Y_d^\dagger Y_d C_{d_R}\right]
                              -2\left[Y_d^\dagger C_{q_L} Y_d - Y_d^\dagger Y_d C_{H}\right]+ y_{d_R} T\mathbb{1}\, , \nonumber\\
(4\pi)^2\frac{dC_{\ell_L} }{d\log\mu} &=& \frac{1}{2} \left[C_{\ell_L} Y_e Y_e^\dagger +  Y_e Y_e^\dagger  C_{\ell_L}\right]
                                  - \left[Y_e C_{e_R} Y_e^\dagger - Y_e Y_e^\dagger C_{H}\right] +y_{\ell_L} T\mathbb{1}\, ,\\
(4\pi)^2\frac{dC_{e_R}}{d\log\mu}  & =&  \left[C_{e_R} Y_e^\dagger Y_e + Y_e^\dagger Y_e C_{e_R}\right]
                             -2 \left[Y_e^\dagger C_{\ell_L} Y_e - Y_e^\dagger Y_e C_{H}\right] +y_{e_R} T\mathbb{1} \, ,\nonumber\\
(4\pi)^2\frac{dC_{H}}{d\log\mu} &=&  2\left(  3\,\mbox{tr}\left[C_{q_L}\hat{Y}^2_q\right]  -3\,\mbox{tr}[Y_u C_{u_R} Y_u^\dagger] \right.
                         +3\, \mbox{tr}[Y_d C_{d_R} Y_d^\dagger] \nonumber\\
                        & & \qquad {} - \mbox{tr}[Y_e^\dagger C_{\ell_L} Y_e]  + \mbox{tr}[Y_e C_{e_R} Y_e^\dagger] \bigg)  +2 \,\mbox{tr}\left[ 3\, Y^2_q+ Y_e^\dagger Y_e \right]C_{H}  + y_H T\, ,\nonumber
\end{eqnarray}
where $Y_\psi$ are non-diagonal Yukawa matrices, and we have defined the useful quantities
\begin{align}\label{eq:useful_aboveEW}
    Y^2_q & \equiv Y_u Y_u^\dagger +Y_d Y_d^\dagger\, , ~~~~ \hat{Y}^2_q \equiv Y_u Y_u^\dag - Y_d Y_d^\dag\, ,
\end{align}
and
\begin{align}\begin{aligned}\label{eq:T}
    T & {} \equiv \frac{4}{3} g'^2\; \bigg(6 y_{q_L} \mathrm{tr}[C_{q_L}] +3  y_{u_R} \mathrm{tr}[C_{u_R}] + 3  y_{d_R} \mathrm{tr}[C_{d_R}] \\
    & \qquad {} + 2  y_{\ell_L} \mathrm{tr}[C_{\ell_L}] +  y_{e_R} \mathrm{tr}[C_{e_R}] + y_H C_H \bigg)\, .
\end{aligned}\end{align}
Notice that all the equations in Eq.~\eqref{eq:RGE_above} contain a term proportional to $T$ generated in the redefinition of the redundant operator. This term is a function of all diagonal elements of the Wilson Coefficients and implies that a coupling between the dark current and a lepton (or quark) current is generated even if not present at the scale $\Lambda$. This fact has been used in the last years to put bounds coming from hadron or lepton collider on leptophilic and leptophobic Dark Matter models~\cite{Crivellin:2014qxa,DEramo:2014nmf,DEramo:2016gos,DEramo:2017zqw}. We have explicitly checked that our results match with those of Ref.~\cite{DEramo:2014nmf} once we restrict to flavor diagonal currents. Let us however remark that the contribution proportional to $T$ is absent in flavor off-diagonal currents. In addition, we expect the largest flavor violating effects to appear on the RGE's involving the top-Yukawa coupling, \textit{i.e.} those of the Wilson coefficients $C_{q_L}^{i3}$, $C_{q_L}^{3i}$, $C_{u_R}^{i3}$ or $C_{u_R}^{3i}$ ($i=1,2$). In the numerical solution of the RGE's (see Section~\ref{sec:results}) we will consider the running of both the gauge and Yukawa couplings at 1-loop as taken from Refs.~\cite{Machacek:1983tz,Machacek:1983fi}. \vv

\begin{figure}[tb]
    \centering
    \includegraphics[height = 0.2 \textwidth]{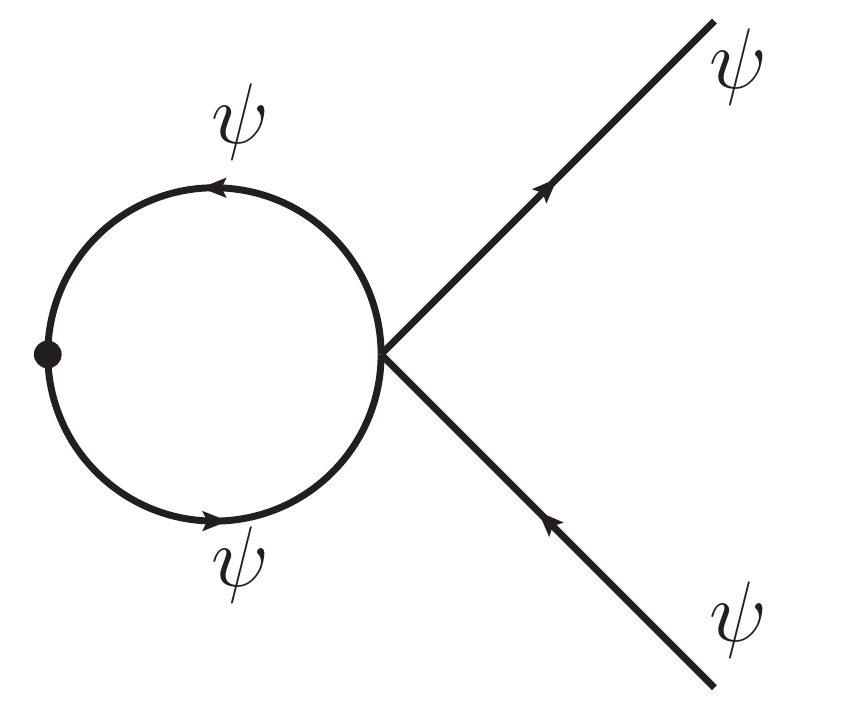} 
    \includegraphics[height = 0.2 \textwidth]{fer_ver_corr_g}
    \includegraphics[height = 0.2 \textwidth]{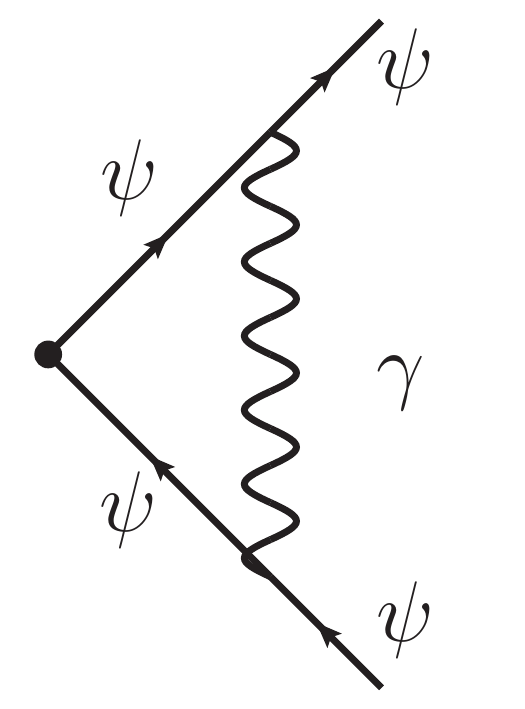}
    \caption{\label{fig:vertex_below} Feynman diagrams contributing to the current renormalization below the EW scale.}
\end{figure}
Let us now turn to the RGE's \textit{below} the EW scale, which we roughly identify with the $Z$ boson mass. At this scale we integrate out the heavy fields $W$, $Z$, $H$ and the $top$ quark. The Lagrangian we consider is
\begin{align}\begin{aligned}\label{eq:EFT_below}
    \mathcal{L} &= \frac{1}{\Lambda^n} J_{\mathcal{D}\mu} \bigg[ \overline{u} \gamma^\mu C_{V_u} u + \overline{d} \gamma^\mu C_{V_d} d + \overline{e} \gamma^\mu C_{V_e} e + \overline{\nu}_L \gamma^\mu C_{V_{L\nu}} \nu_L \\ 
                & \qquad {} + \overline{u} \gamma^\mu \gamma_5 C_{A_u} u + \overline{d} \gamma^\mu \gamma_5 C_{A_d} d + \overline{e} \gamma^\mu \gamma_5 C_{A_e} e \bigg]\, ,
\end{aligned}\end{align}
where $C_{V_u}$ and $C_{A_u}$ are $2\times 2$ matrices in the up-type quark flavor space, while all the other matrices are $3\times 3$ in flavor space. Notice that we are not considering the current $\partial_\nu F^{\nu\mu}$ since it can be redefined away using the photon equations of motion. The procedure to obtain the RGE's is as before. The corrections due to the wave function renormalization are now due only to strong and electromagnetic interactions, and again they cancel against the vertex corrections in Fig.~\ref{fig:vertex_below}. What remains of the vertex corrections are the Fermi contributions, with flavor diagonal and off-diagonal contributions coming from neutral and charged current interactions, respectively. Again, we need to take care of the redundant current $\partial_\nu F^{\nu\mu}$ which is generated radiatively by diagrams similar to those shown in Fig.~\ref{fig:redop_corr} (with a photon in the external leg instead of a $B$ vector), eliminating it via the EoM of the photon field. Again, more details are shown in Appendix~\ref{app:computation_RGE}. The RGE's are given by

    \begin{eqnarray}\label{eq:RGE_below}
    (4\pi)^2\frac{d C_{V_u}}{d \log \mu} &=& g_{V_u}
    \mathcal{F}_{u}+{G_F}_{ud}  V \big( \mathcal{M}^2_{V_d}- \mathcal{M}^2_{A_d}\big) V^\dag + Q_u  \mathcal{Q} , \nonumber
    \\
    (4\pi)^2\frac{d C_{V_d}}{d \log \mu} & = &  g_{V_d}
    \mathcal{F}_{d}+{G_F}_{du}   V^\dag \big(\mathcal{M}^2_{V_u} -  \mathcal{M}^2_{A_u} \big) V + Q_d  \mathcal{Q},  \nonumber
    \\
    (4\pi)^2\frac{d C_{V_e}}{d \log \mu} & =&   g_{V_e}
    \mathcal{F}_{e}+  Q_e \mathcal{Q},  \nonumber
    \\
    (4\pi)^2\frac{d C_{V_{L\nu}}}{d \log \mu}  &=&  g_{V_\nu}
    \mathcal{F}_{\nu}+{G_F}_{\nu e}\left(\mathcal{M}^2_{V_e} -  \mathcal{M}^2_{A_e}\right),
    \\
    (4\pi)^2\frac{d C_{A_u}}{d \log \mu} & =& g_{A_u}
    \mathcal{F}_{u}    - {G_F}_{ud}V \big( \mathcal{M}^2_{V_d}- \mathcal{M}^2_{A_d}\big) V^\dag,  \nonumber
    \\
    (4\pi)^2\frac{d C_{A_d}}{d \log \mu}  &=&  g_{A_d}
    \mathcal{F}_{d}- {G_F}_{du} V^\dag \big(\mathcal{M}^2_{V_u} -  \mathcal{M}^2_{A_u} \big) V,  \nonumber
    \\
    (4\pi)^2\frac{d C_{A_e}}{d \log \mu} &=&  g_{A_e}
    \mathcal{F}_{e},  \nonumber
    \end{eqnarray}
    where $V$ is the CKM matrix, and we have used the definitions  
\begin{align}\begin{aligned}
    \mathcal{M}^2_{V_i} &= 2\sqrt{2}\left(C_{V_i} M_i^2 + M_i^2 C_{V_i} -2 M_i C_{V_i} M_i\right)\, , \\
    \mathcal{M}^2_{A_i} &= 2\sqrt{2}\left(C_{A_i} M_i^2 + M_i^2 C_{A_i} +2 M_i C_{A_i} M_i\right)\, . 
\end{aligned}\end{align}
The matrix $M_i$ is the diagonal mass matrix for the fermions of type $i$, and we have defined
\begin{align}\begin{aligned}\label{eq:useful_below}
    \mathcal{Q} & = \frac{8}{3}e^2\big[ 3 Q_u \mbox{tr}[C_{V_u}] + 3 Q_d \mbox{tr}[C_{V_d}] + Q_e \mbox{tr}[C_{V_e}] \big] \mathbb{1}\, , \\
\mathcal{F}_{i} &= 16\sqrt{2}\big[3 g_{A_u} {G_F}_{iu}\mbox{tr}[M_{u}^2 C_{A_u}]  \\
                & \qquad {} + 3 g_{A_d} {G_F}_{id} \mbox{tr}[M_{d}^2 C_{A_d}] + g_{A_e} {G_F}_{ie} \mbox{tr}[M_{e}^2 C_{A_e}]\big]  \mathbb{1}.
\end{aligned}\end{align}
The coefficients $g_{V_i}=T^3_i-2s^2_w Q_i$ and $g_{A_i}=-T^3_i$ are the vector and axial couplings ($T^3_i$ is the third component of the isospin, $Q_i$ is the electric charge and $s_w$ is the sine of the weak angle), while the Fermi couplings ${G_F}_{ff'}$ are defined in Eqs.~\eqref{eq:neutralcurr} and~\eqref{eq:chargedcurr}. It should be noted that, below the EW scale, not only the fermion masses run, but we need also to take into account the running of ${G_F}_{ff'}$, and the running depends on the fermion type $f$ and $f'$ involved, justifying in this way the fact that we do not write a unique Fermi coupling. We use Ref.~\cite{Xing:2011aa} for the running of the masses, and show in Appendix~\ref{app:running_Gf} more details on the running of ${G_F}_{ff'}$. \vv

Before closing this Section, let us comment on how the two theories match onto each other. More specifically, when the scale $\Lambda$ at which the dark sector EFT is generated is above the EW scale, the operators of Eq.~\eqref{eq:EFT_above} run and mix according to Eq.~\eqref{eq:RGE_above} down to $\Lambda_{EW} \simeq m_Z$. At this scale, the operators must be matched onto the Lagrangian of Eq.~\eqref{eq:EFT_below} before continuing with the running of Eq.~\eqref{eq:RGE_below} down to low energies. This procedure was presented in detail in  Ref.~\cite{DEramo:2014nmf} for flavor diagonal Wilson coefficients. In our case, the only new feature is that once we cross the EW threshold, all the fermions must be rotated into the mass basis. Explicitly, we write this transformation as
\begin{align}
 	f_L \to L_f f_L\, , ~~~~~ f_R \to R_f f_R\, ,
\end{align}
where the matrices $L_f$ and $R_f$ diagonalize the Yukawa matrices. The matching then results
\begin{align}\begin{aligned}\label{eq:matching}
C_{V_u}(\Lambda_{EW}) & = \frac{1}{2}\bigg(L_{q}^\dag C_{q_L}(\Lambda_{EW})L_{q} + R_u^\dag C_{u_R} (\Lambda_{EW})R_u \bigg) + g_{V_u} \, C_H(\Lambda_{EW}) \mathbb{1}\, ,\\
C_{V_d}(\Lambda_{EW}) & = \frac{1}{2}\bigg(L_q^\dag C_{q_L}(\Lambda_{EW}) L_q + R_d^\dag C_{d_R} (\Lambda_{EW})R_d \bigg) + g_{V_d} \, C_H(\Lambda_{EW}) \mathbb{1}\, ,\\
C_{V_{L\nu}}(\Lambda_{EW}) & = \frac{1}{2} L_\ell^\dag C_{\ell_L}(\Lambda_{EW}) L_\ell+ g_{V_\nu} \, C_H(\Lambda_{EW}) \mathbb{1}\, ,\\
C_{V_e}(\Lambda_{EW}) & = \frac{1}{2}\bigg(L_\ell^\dag C_{\ell_L}(\Lambda_{EW}) L_\ell + R_e^\dag C_{e_R} (\Lambda_{EW}) R_e\bigg) + g_{V_e} \, C_H(\Lambda_{EW}) \mathbb{1}\, ,\\
C_{A_u}(\Lambda_{EW}) & = \frac{1}{2}\bigg(-L_q^\dag C_{q_L}(\Lambda_{EW}) L_q + R_u^\dag C_{u_R}(\Lambda_{EW}) R_u\bigg) + g_{A_u} \, C_H(\Lambda_{EW}) \mathbb{1}\, ,\\
C_{A_d}(\Lambda_{EW}) & = \frac{1}{2}\bigg(-L_q^\dag C_{q_L}(\Lambda_{EW}) L_q + R_d^\dag C_{d_R}(\Lambda_{EW}) R_d\bigg) + g_{A_d} \, C_H(\Lambda_{EW}) \mathbb{1}\, ,\\
%C_{A_\nu}(\Lambda_{EW}) & = -\frac{1}{2} C_{\ell_L}(\Lambda_{EW}) + g_{A_\nu} \, C_H(\Lambda_{EW}) \mathbb{1}\, ,\\
C_{A_e}(\Lambda_{EW}) & = \frac{1}{2}\bigg(-L_\ell^\dag C_{\ell_L}(\Lambda_{EW}) L_\ell + R_e^\dag C_{e_R}(\Lambda_{EW}) R_e \bigg) + g_{A_e} \, C_H(\Lambda_{EW}) \mathbb{1}\, .\\
\end{aligned}\end{align}
Notice in particular that the term induced by $C_H$ affects only the diagonal Wilson coefficients. In the following Section we will solve numerically the RGE's and show the numerical impact of turning on off-diagonal currents at the scale $\Lambda$.

%%%%%%%%%%%%%%%%%%%%%%%%%%%
\section{Numerical results and possible applications}\label{sec:results}
%%%%%%%%%%%%%%%%%%%%%%%%%%%

\begin{figure}[tb]
	\centering
	\includegraphics[width = 0.49 \textwidth]{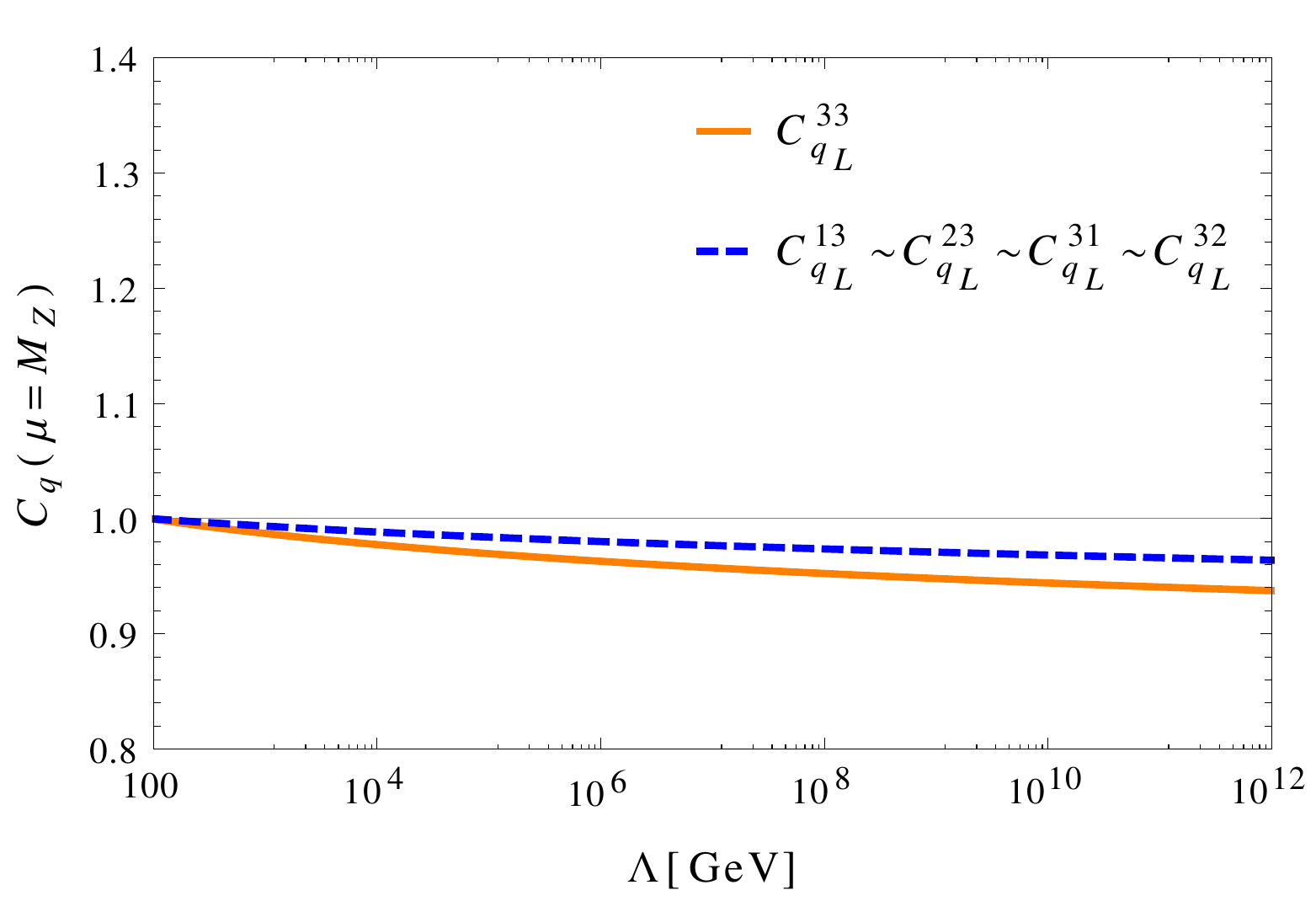}
	 \includegraphics[width = 0.49 \textwidth]{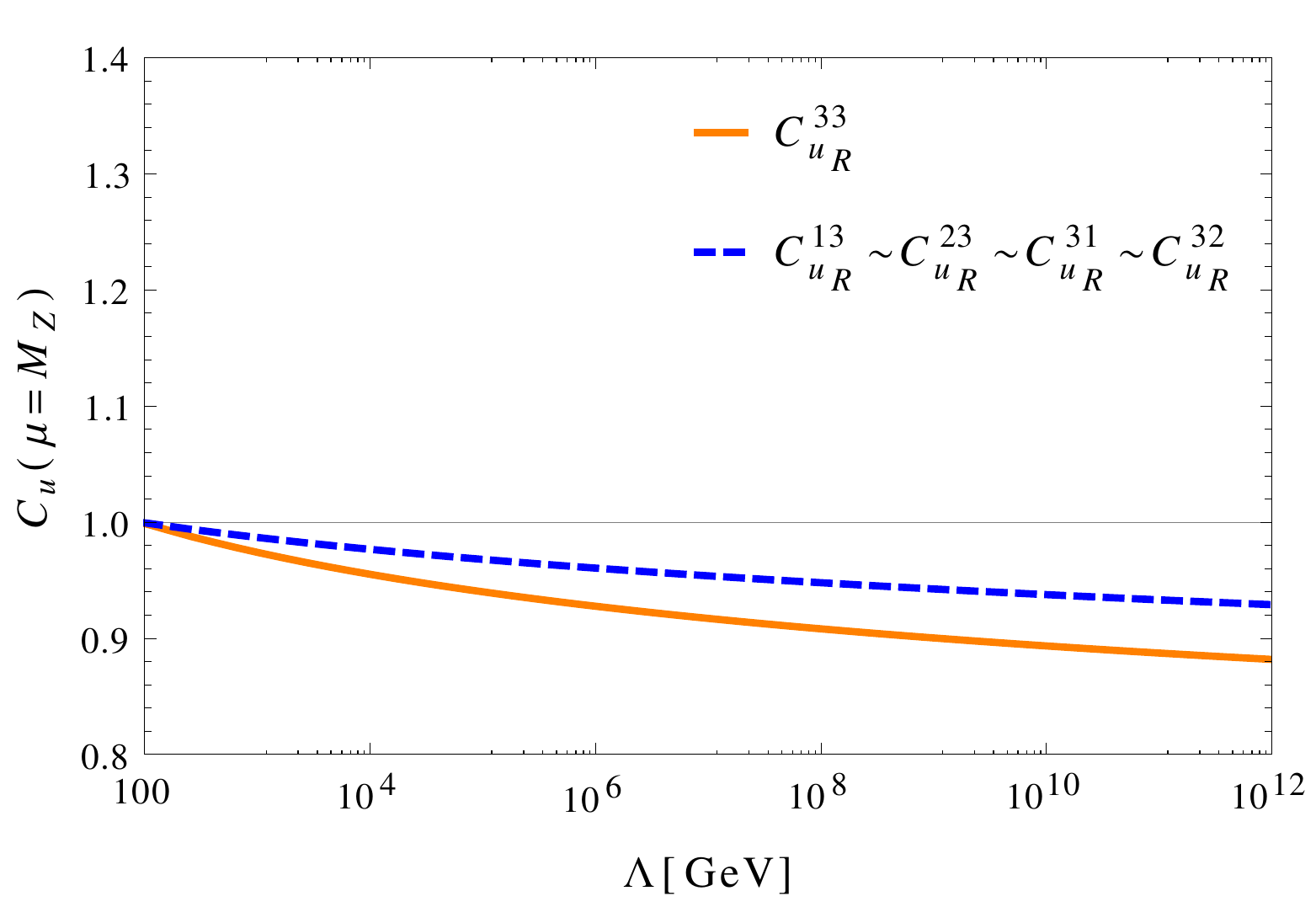}\\
	 \includegraphics[width = 0.49 \textwidth]{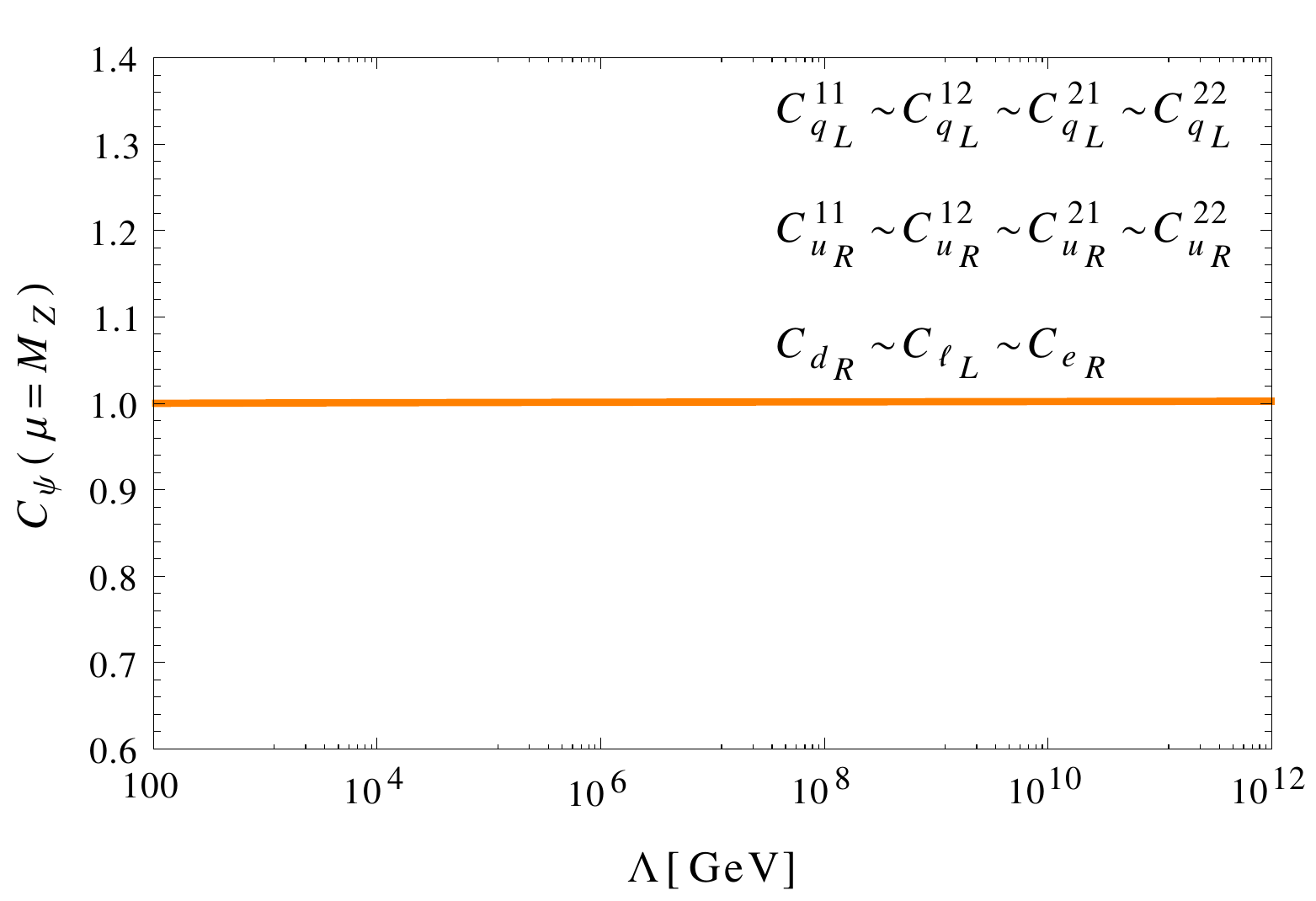}
	 \includegraphics[width = 0.49 \textwidth]{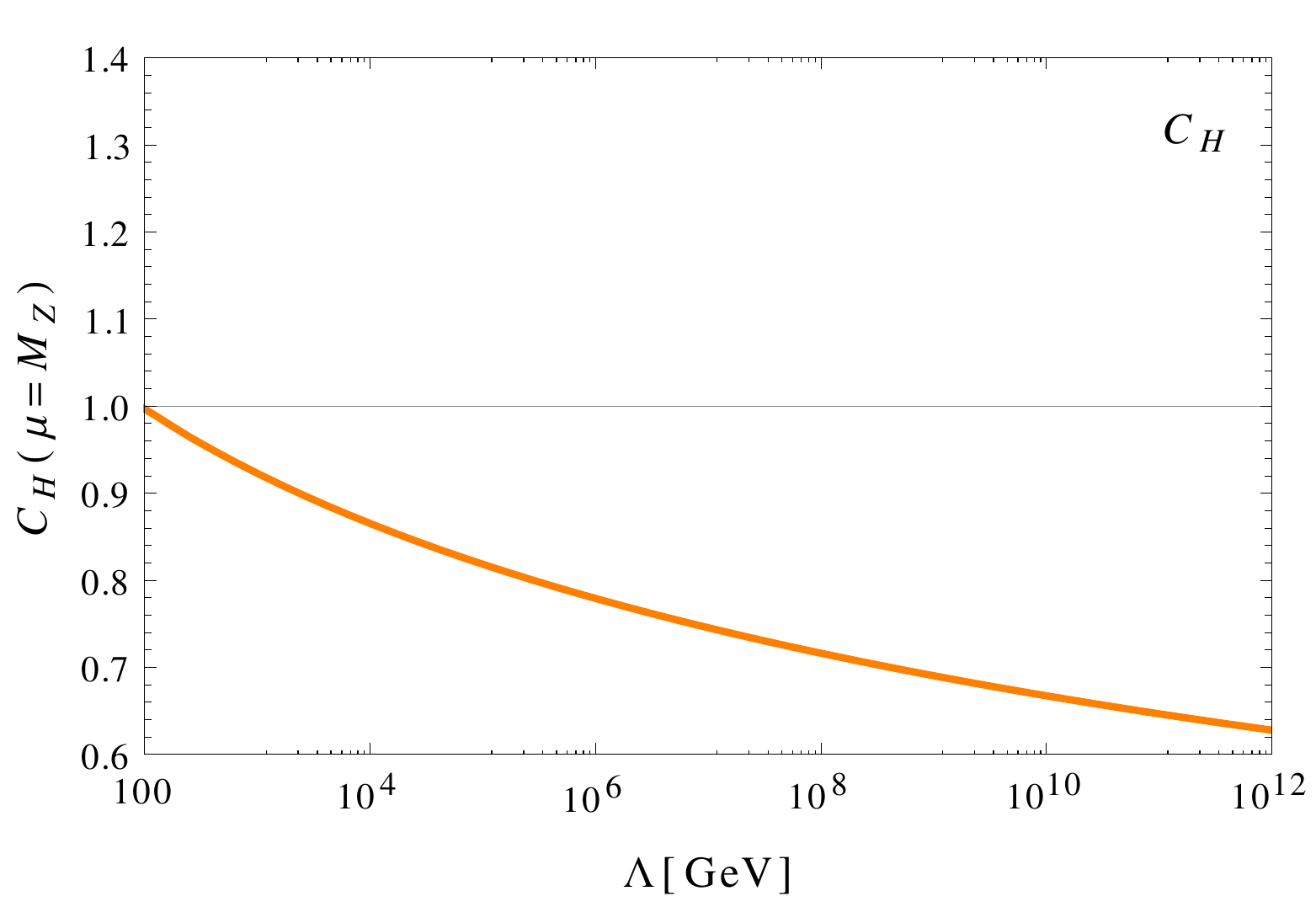}
	\caption{\label{fig:cAllabove} Running of the Wilson Coefficients above the EW scale, taking as initial condition $(C_a)_{ij}(\Lambda) = 1$ for all the values of $a$, $i$ and $j$. In the bottom left panel, $C_{d_R} \sim C_{\ell_L} \sim C_{e_R}$ applies to all the matrix elements, with no difference for those involving the third generation.}
\end{figure}

We turn in this section to the numerical solution of the RGE's presented in Eqs.~\eqref{eq:RGE_above} and~\eqref{eq:RGE_below}. As already mentioned, in solving Eqs.~\eqref{eq:RGE_above} and~\eqref{eq:RGE_below} we not only take into account the EW threshold if needed, but we also consider the running of the gauge and Yukawa couplings (for the theory above the EW scale) and the running of the Fermi coupling and fermion masses (for the theory below the EW scale). We show in Fig.~\ref{fig:cAllabove} the value of the Wilson coefficients at the EW scale as a function of $\Lambda$, considering the ``flavor-democratic'' initial condition
\begin{align}
    (C_a)_{ij}(\Lambda) = 1\, , ~~~~~ \mathrm{for\;all\;}a, i\;\mathrm{and}\;j\, .
\end{align}
As expected, the effect of the running is more important for the currents constructed out of top-quarks (with Wilson coefficients $C_{q_L}^{3i}$, $C_{q_L}^{i3}$, $C_{u_R}^{3i}$ and $C_{u_R}^{i3}$, with $i=1,2,3$) and for $C_H$, in which the top Yukawa coupling enters. The result can be easily understood inspecting Eq.~\eqref{eq:RGE_above}. In the limit in which only the top-Yukawa coupling is turned on, we have
\begin{align}\begin{aligned}\label{eq:approx_RGE_above}
    \frac{dC_{q_L}^{i3}}{d\log\mu} & \simeq  \frac{y_t^2}{32\pi^2} C_{q_L}^{i3} \,, & \frac{dC_{q_L}^{3i}}{d\log\mu} & \simeq  \frac{y_t^2}{32\pi^2} C_{q_L}^{3i} \,, & \frac{dC_{q_L}^{33}}{d\log\mu} & \simeq  \frac{y_t^2}{16\pi^2} \left(C_H + C_{q_L}^{33} - C_{u_R}^{33} \right)\,,  \\
    \frac{dC_{u_R}^{i3}}{d\log\mu} & \simeq  \frac{y_t^2}{16\pi^2} C_{u_R}^{i3} \,,  & \frac{dC_{u_R}^{3i}}{d\log\mu} & \simeq  \frac{y_t^2}{16\pi^2} C_{u_R}^{3i} \,, & \frac{dC_{u_R}^{33}}{d\log\mu} & \simeq  \frac{y_t^2}{8\pi^2} \left(C_H - C_{q_L}^{33} + C_{u_R}^{33} \right)\,,  \\
\end{aligned}\end{align}
with $i=1,2$ and all the other RGE's vanishing. The solutions at the scale $m_t$ at which we integrate out the top quark are easily found. For the off-diagonal Wilson coefficients they are
\begin{align}\begin{aligned}\label{eq:sol_approx_RGE_offdiag}
    C_{q_L}^{i3}(m_t) &\simeq \left( \frac{m_t}{\Lambda} \right)^{y_t^2/32\pi^2} C_{q_L}^{i3}(\Lambda) \, , & ~ & C_{q_L}^{3i}(m_t) &\simeq \left( \frac{m_t}{\Lambda} \right)^{y_t^2/32\pi^2} C_{q_L}^{3i}(\Lambda) \, , \\
    C_{u_R}^{i3}(m_t) &\simeq \left( \frac{m_t}{\Lambda} \right)^{y_t^2/16\pi^2} C_{u_R}^{i3}(\Lambda) \, , & ~ & C_{u_R}^{3i}(m_t) &\simeq \left( \frac{m_t}{\Lambda} \right)^{y_t^2/16\pi^2} C_{u_R}^{3i}(\Lambda) \, , 
\end{aligned}\end{align}
where again $i=1,2$, while for the diagonal Wilson coefficients they are
\begin{align}\begin{aligned}\label{eq:sol_approx_RGE_diag}
    C_{q_L}^{33}(m_t) &\simeq C_{q_L}^{33}(\Lambda) + \frac{y_t^2}{16\pi^2} \left[C_H(\Lambda) + C_{q_L}^{33}(\Lambda) - C_{u_R}^{33}(\Lambda)  \right] \log\frac{m_t}{\Lambda}  \, , \\ C_{u_R}^{33}(m_t) &\simeq C_{u_R}^{33}(\Lambda) + \frac{y_t^2}{8\pi^2} \left[C_H(\Lambda) - C_{q_L}^{33}(\Lambda) + C_{u_R}^{33}(\Lambda)  \right] \log\frac{m_t}{\Lambda}\, . \\
\end{aligned}\end{align}

From these equations we get various important informations: (i) the off-diagonal Wilson coefficients can have a substantial running due to $y_t$, (ii) the renormalization of the flavor-off-diagonal Wilson coefficients is always multiplicative, in the sense that there are no important flavor changing generated during the running to low energy, and (iii) while the running of the flavor-off-diagonal Wilson coefficients is irreducible, the running of the flavor-diagonal coefficients depends strongly on the correlations between the initial conditions of different Wilson coefficients. For instance, we see from Eq.~\eqref{eq:sol_approx_RGE_diag} that had we considered ``flavor-democratic'' initial conditions without generating the Higgs current ($C_H(\Lambda)=0$), then the diagonal Wilson coefficients would basically not run. This is not true for the off-diagonal coefficients, that once turned on will run independently from the initial conditions of other coefficients. We stress that the analytical expressions of Eqs.~(\ref{eq:sol_approx_RGE_offdiag}) and (\ref{eq:sol_approx_RGE_diag}) reproduce accurately the full numerical results.\vv

Moving to the EFT below the EW scale, inspection of Eq.~\eqref{eq:RGE_below} shows that the RGE's depend on the Fermi coupling ${G_F}_{ff'}$ and on the electric charge $e^2$. Both contributions are rather small, and from the practical point of view, all the Wilson coefficients remain basically constant in this energy range. We have confirmed numerically that this is indeed the case. We conclude then that the relevant running happens above $m_t$, in the energy region in which the top-quark is still a dynamical degree of freedom. For mediators with mass above the EW scale coupling to the top quark, the effect of the running may be important and must thus be taken into account in the comparison with experiments. On the contrary, if the mediator is lighter than the EW scale, or if it does not interact with the top quark, the tree level predictions are usually a good approximation for the extractions of phenomenological bounds.\vv

We conclude this Section with some phenomenological remark. Using Eq.~\eqref{eq:sol_approx_RGE_offdiag} we see that the main effect of the running will manifest at low energy in observables related to the $B$ mesons, \textit{i.e.} involving the $\overline{b} \gamma^\mu (\gamma_5)d$ and $\overline{b} \gamma^\mu (\gamma_5) s$ currents. More precisely, the vector and axial coefficients at a scale $\mu \ll m_Z$ are given by
\begin{align}\begin{aligned}
    C_{V_d}^{3i}(\mu) &\simeq \frac{1}{2} \left[\left( \frac{m_t}{\Lambda} \right)^{y_t^2/32\pi^2} C_{q_L}^{i3}(\Lambda) + C_{d_R}^{3i}(\Lambda) \right]\, , \\
    C_{A_d}^{3i}(\mu) &\simeq \frac{1}{2} \left[-\left( \frac{m_t}{\Lambda} \right)^{y_t^2/32\pi^2} C_{q_L}^{i3}(\Lambda)  + C_{d_R}^{3i}(\Lambda) \right]\, , \\
\end{aligned}\end{align}
the same result is also valid by changing the order $3 \leftrightarrow i$, with $i=1,2$. This is important when the dark sector is light, in such a way that the decays $b \to s + \mathrm{invisible}$ and $b \to d + \mathrm{invisible}$ are kinematically allowed. These processes were studied in the context of dark sector phenomenology in Ref.~\cite{Kamenik:2011vy}. As shown in this Reference, the bounds on the Wilson coefficients depend crucially on the nature of the dark particles appearing in the dark current. Given this model dependence, we will not explore this matter here. In addition, we remark that more flavor effects would be obtained once dark particles loops are considered, for instance generating contributions to the $B$ mesons mass mixing. As shown in Ref.~\cite{Bishara:2018vix} in the context of flavor diagonal DM EFT, these effects can be important, but since they are model dependent, we defer their study to a future work.

%%%%%%%%%%%%%%%%%%%%%%%%%%%
\section{Summary and conclusions}\label{sec:summary}
%%%%%%%%%%%%%%%%%%%%%%%%%%%

In this paper we have studied the running and mixing of operators of a dark sector EFT, under the hypothesis that the operators are generated by a heavy \textit{flavorful} mediator. We have carefully computed the running above and below the EW scale and the matching due to the heavy SM particles. Our main results are shown in Fig.~\ref{fig:cAllabove} and in the approximate analytic solutions of Eqs.~\eqref{eq:sol_approx_RGE_offdiag} and~\eqref{eq:sol_approx_RGE_diag}. The most important effects in the running are generated by the top Yukawa coupling, and as such are present only above the top quark threshold. Once turned on at the scale $\Lambda$, the contribution of the running on the flavor-off-diagonal Wilson coefficients is irreducible, in the sense that, unlike what happens for flavor-diagonal coefficients, it does not depend on possible correlations between the Wilson coefficients at the scale $\Lambda$. From a phenomenological perspective, our results imply that the most important effect of the running and mixing is found at low energy in the currents constructed out of the left-handed down-type quarks $\overline{b}_L \gamma^\mu d_L$ and $\overline{b}_L \gamma^\mu  s_L$. If the dark sector is heavier than the $B$ mesons mass scale, then the main effects will be due to loops of dark particles. If instead the dark sector is sufficiently light, the decays $b \to s + \mathrm{invisible}$ and $b \to d + \mathrm{invisible}$ may be kinematically allowed and can be used to put bounds on the Wilson coefficients. Since all these processes depend on the specific nature of the dark sector particles, we defer their study to a forthcoming publication. 

\acknowledgments

We thank Renata Zukanovich Funchal and Oscar Éboli for comments on the manuscript. E.B. would like to thank Francesco D'Eramo for useful conversations. Z.T. appreciates the useful discussions with Yuber Perez. This work was supported by Funda\c{c}\~{a}o de Amparo \`{a} Pesquisa (FAPESP), under contracts 16/02636-8 and 15/25884-4, and Conselho Nacional de Ci\^{e}ncia e Tecnologia (CNPq). 

\appendix

\section{Standard Model: conventions}\label{app:details}

We show in this Appendix a few more details on the SM Lagrangian, as well as some useful definitions used throughout the paper. According to the conventions shown in Table~\ref{field_rep}, the SM Lagrangian in the unbroken phase is given by
\begin{align}\begin{aligned}\label{eq:lagr_SM_above}
	\mathcal{L} =& -\frac{1}{4} (G_{\mu\nu})^2 - \frac{1}{4} (W^a_{\mu\nu})^2 - \frac{1}{4} (B_{\mu\nu})^2 \\
				& + i\overline{\psi} \slashed{D}\psi + \left|D_\mu H \right|^2 - V(H) \\
				& + \overline{q}_L Y_d H d_R + \overline{q}_L Y_u \tilde{H} u_R + \overline{\ell}_L Y_e H e_R + h.c.\, ,
\end{aligned}\end{align}
where as usual $\tilde{H} = i \sigma_2 H^*$. We stress that we do not commit to a fermion basis in which some of the Yukawa matrices $Y_u$, $Y_d$ and $Y_e$ are diagonal. The equation of motion of the hypercharge vector field is particularly important, since it allows to define away a redundant current. It reads
\begin{align}\label{eq:EOM_1}\begin{aligned}
	\partial^\nu B_{\mu\nu} = g' J_\mu^Y =& g' \left(y_{q_L} \overline{q}_L \gamma_\mu q_L + y_{u_R} \overline{u}_R\gamma_\mu u_R + \overline{d}_R\gamma_\mu d_R + y_{\ell_L} \overline{\ell}_L \gamma_\mu \ell_L  \right. \\
	& \left. y_{e_R} \overline{e}_R \gamma_\mu e_R + y_H i H^\dag \overleftrightarrow{D}_\mu H \right)\, .
\end{aligned}\end{align}
Once the EW symmetry is broken by the Higgs vacuum expectation value (vev)
\begin{align}
	\langle H \rangle = \begin{pmatrix} 0 \\ v \end{pmatrix}\, ~~~~~ v \simeq 174\; \mathrm{GeV}\, ,
\end{align}
and all the states get non-vanishing masses, we integrate out the heavy fields. The SM Lagrangian in the broken phase (\textit{i.e.} below the EW scale) results
\begin{align}\begin{aligned}
	\mathcal{L} &= -\frac{1}{4} (G_{\mu\nu})^2 - \frac{1}{4} (F_{\mu\nu})^2  + \overline{\psi} (\slashed{D}-m_\psi) \psi + \mathcal{L}_{\rm Fermi}\, ,\\
\end{aligned}\end{align}
where $F_{\mu\nu}$ is the photon field strength, while $\psi$ represents any of the light Dirac fermions still present in the theory. We do not assume the existence of right handed neutrinos in the low energy spectrum, and we leave unspecified the mechanism behind their mass generation. Once the $Z$ and $W$ bosons are integrated out, we obtain two contributions for the four-fermion Fermi Lagrangian. The neutral current one reads
\begin{align}\label{eq:neutralcurr}
\mathcal{L}_{\rm F} \supset -\frac{{G_F}_{ff'}}{\sqrt{2}} g_{\Gamma f} g_{\Gamma' f'} (\overline{f} \Gamma^\mu f) (\overline{f'} \Gamma'_\mu f')\, ,
\end{align}
where $\Gamma = \gamma^\mu$ or $\gamma^\mu \gamma_5$ as defined below Eq.~\eqref{eq:useful_below}. Although the notation ${G_F}_{ff'}$ seems redundant, we keep explicit track of the `fermion indices' because, as we will show in Appendix~\ref{app:running_Gf}, different $f$ and $f'$ correspond to different running for the coupling. 

Turning to the charged current, the relevant contributions read
\begin{align}\label{eq:chargedcurr}
\mathcal{L}_{\rm F} \supset -\frac{{G_F}_{ud}}{4\sqrt{2}} \sum _{i,j,k,l} V^{ij}V^{\dagger kl}  \bar{u}^i \left(\gamma^{\mu }-\gamma^{\mu}\gamma_5 \right) u^l \bar{d}^k \left(\gamma_{\mu }-\gamma_{\mu }\gamma_5 \right) d^j,
\end{align}
where $V$ denotes the CKM matrix.

The EoM of the photon field, used to eliminate the redundant operator $\partial_\mu F^{\nu\mu}$ is
\begin{align}
	\partial_\mu F^{\nu\mu} = e J_{em}^\nu\, ,
\end{align}
with $J_{em}$ the usual electromagnetic current.

\section{EFT generated in specific models}\label{app:models}

Let us now show how some of the flavor models involving dark sectors present in the literature can be mapped to our formalism. We start with the situation considered in References~\cite{Agrawal:2014aoa,Blanke:2017fum,Blanke:2017tnb}, which can be summarized via the interactions
\begin{align}
	\mathcal{L}_{int}^A = \lambda_{ij} \overline{d}_R^i \chi_L^j \phi\, , ~~~ \mathcal{L}_{int}^B = \lambda_{ij} \overline{u}_R^i \chi_L^j \phi\, , ~~~ \mathcal{L}_{int}^C = \lambda_{ij} \overline{q}_L^i \chi_R^j \phi\, .
\end{align}
The triplet of dark fermions $\chi_i$ is chosen to be a gauge singlet, while the scalar mediator $\phi$ transforms under the SM gauge group as the quark to which it couples to. The same kind of interactions have been considered in Ref.~\cite{Beauchesne:2017yhh} in the context of collider searches of Hidden Valley models. Taking the mediator to be heavy, we obtain at low energy
\begin{align}\begin{aligned}\label{eq:explicit_EFT}
	\mathcal{L}_{EFT}^A &= \frac{\lambda_{ij} \lambda_{km}^*}{2 m_\phi^2} \big(\overline{\chi}_L^k \gamma^\mu \chi_L^j\big)  \big(\overline{d}_{R}^i \gamma_\mu d_R^m  \big)\, , \\
	\mathcal{L}_{EFT}^B &= \frac{\lambda_{ij} \lambda_{km}^*}{2 m_\phi^2} \big(\overline{\chi}_L^k \gamma^\mu \chi_L^j \big) \big(\overline{u}_{R}^i \gamma_\mu u_R^m  \big)\, , \\
	\mathcal{L}_{EFT}^C &= \frac{\lambda_{ij} \lambda_{km}^*}{2 m_\phi^2} \big(\overline{\chi}_R^k \gamma^\mu \chi_R^j \big) \big(\overline{q}_{L}^i \gamma_\mu q_L^m  \big) \, .\\
\end{aligned}\end{align}
This shows that the combination of operators studied in this paper can be easily obtained in specific models.

\section{Computation of the Renormalization Group Equation}\label{app:computation_RGE}

In this appendix we will present more details on the computation of the RGE's for the Wilson coefficients of the $d=3$ currents appearing in Tables~\ref{tab:SM_currents_aboveEW} and~\ref{tab:SM_currents_belowEW}. The final results have been shown in Sec.~\ref{sec:RGEs}, with numerical solutions given in Sec.~\ref{sec:results}. We will always use dimensional regularization in $d = 4-2\varepsilon$ dimensions, and use the $\overline{\mathrm{MS}}$ scheme.

Let us start by considering loop corrections \textit{above the EW scale}. The counterterm Lagrangian $\mathcal{L}_{c.t.}$ generated by the wave function renormalization (see  Fig.~\ref{fig:wave_function}) is given by
\begin{align}\begin{aligned}
	\small
	\mathcal{L}_{c.t.}  & =  \left[\frac{g^2 \,\mathcal{C}(\mathbf{2}) + g'^2 \,y_H^2}{8\pi^2 \varepsilon} - \frac{3\, \mathrm{tr}(Y_q^2)+  \mathrm{tr}(Y_e^\dag Y_e)}{16\pi^2 \varepsilon} \right] \partial_\mu H^\dag \partial^\mu H  \\
	&  {} - \overline{q}_L  \left[\frac{g_s^2\, \mathcal{C}(\mathbf{3}) + g^2\, \mathcal{C}(\mathbf{2}) + g'^2\, y_{q_L}^2 }{16\pi^2 \varepsilon} \mathbb{1} + \frac{Y_q^2}{32\pi^2 \varepsilon} \right] i \slashed{\partial} q_L \\
	&  {}- \overline{u}_R \left[  \frac{g_s^2\, \mathcal{C}(\mathbf{3}) + g'^2\, y_{u_R}^2 }{16\pi^2 \varepsilon} \mathbb{1} + \frac{Y_u^\dag Y_u}{16\pi^2 \varepsilon} \right] i \slashed{\partial} u_R - \overline{d}_R \left[\frac{g_s^2\, \mathcal{C}(\mathbf{3}) +  g'^2\, y_{d_R}^2 }{16\pi^2 \varepsilon} \mathbb{1} + \frac{Y_d^\dag Y_d}{16\pi^2 \varepsilon}   \right] i \slashed{\partial} d_R \\
	&  {} - \overline{\ell}_L \left[\frac{ g^2\, \mathcal{C}(\mathbf{2}) + g'^2\, y_{\ell_L}^2 }{16\pi^2 \varepsilon} \mathbb{1} + \frac{Y_e Y_e^\dag }{32\pi^2 \varepsilon}  \right] i \slashed{\partial} \ell_L - \overline{e}_R \left[\frac{ g'^2\, y_{e_R}^2 }{16\pi^2 \varepsilon} \mathbb{1} + \frac{Y_e^\dag Y_e }{16\pi^2 \varepsilon}   \right] i \slashed{\partial} e_R\, ,
\end{aligned}\end{align} 
where $\mathcal{C}(\mathbf{2})$ and $\mathcal{C}(\mathbf{3})$ are, respectively, the $SU(2)_L$ and $SU(3)_c$ Casimirs for the fundamental representations, $y_i$ denotes the field hypercharge and we have used Eq.~\eqref{eq:useful_aboveEW} for the definition of $Y_q^2$. The connection between renormalized and bare fields is now straightforwardly found:
\begin{align}\begin{aligned}\label{eq:wavefunction}
	H & \simeq \left[ \mathbb{1} - \frac{1}{2} \left( \frac{g^2 \,\mathcal{C}(\mathbf{2}) + g'^2 \,y_H^2}{8\pi^2 \varepsilon} - \frac{3\, \mathrm{tr}(Y_q^2)+  \mathrm{tr}(Y_e^\dag Y_e)}{16\pi^2 \varepsilon}  \right)\right]H_{bare} \, , \\
	q_L &\simeq \left[\mathbb{1} + \frac{1}{2}\left( \frac{g_s^2\, \mathcal{C}(\mathbf{3}) + g^2\, \mathcal{C}(\mathbf{2}) + g'^2\, y_{q_L}^2 }{16\pi^2 \varepsilon} \mathbb{1} + \frac{Y_q^2}{32\pi^2 \varepsilon} \right)\right]q_{L,bare}\, , \\
	u_R &\simeq \left[\mathbb{1} + \frac{1}{2}\left(\frac{g_s^2\, \mathcal{C}(\mathbf{3}) + g'^2\, y_{u_R}^2 }{16\pi^2 \varepsilon} \mathbb{1} + \frac{Y_u^\dag Y_u}{16\pi^2 \varepsilon} \right)\right] u_{R,bare}\, , \\
	d_R &\simeq \left[\mathbb{1} + \frac{1}{2}\left(\frac{g_s^2\, \mathcal{C}(\mathbf{3}) +  g'^2\, y_{d_R}^2 }{16\pi^2 \varepsilon} \mathbb{1} + \frac{Y_d^\dag Y_d}{16\pi^2 \varepsilon} \right)\right] d_{R,bare} \, ,\\
	\ell_L & \simeq \left[\mathbb{1} + \frac{1}{2}\left(\frac{ g^2\, \mathcal{C}(\mathbf{2}) + g'^2\, y_{\ell_L}^2 }{16\pi^2 \varepsilon} \mathbb{1} + \frac{Y_e Y_e^\dag }{32\pi^2 \varepsilon} \right)\right] \ell_{L,bare} \, ,\\
	e_R &\simeq \left[\mathbb{1} + \frac{1}{2}\left(\frac{ g'^2\, y_{e_R}^2 }{16\pi^2 \varepsilon} \mathbb{1} + \frac{Y_e^\dag Y_e }{16\pi^2 \varepsilon} \right)\right] e_{R,bare}\, . \\
\end{aligned}\end{align}
Our results confirm the computation of Ref.~\cite{DEramo:2014nmf}. Let us now turn to the computation of the counterterms due to currents corrections (see Fig.~\ref{fig:ver_corr}). For each Wilson coefficient $C_a$, direct computation gives the counterterms
\begin{align}\begin{aligned}\label{eq:ct_vertex_above_mZ}
	\small
	\delta C_{q_L} =& - \frac{g_s^2\, \mathcal{C}(\mathbf{3}) + g^2\, \mathcal{C}(\mathbf{3}) + g'^2 \, y_{q_L}^2}{16\pi^2 \varepsilon}C_{q_L} - \frac{Y_u C_{u_R} Y_u^\dag + Y_d C_{d_R} Y_d^\dag}{32\pi^2 \varepsilon} + \frac{Y_q^2}{32\pi^2 \varepsilon} C_H\, , \\
	\delta C_{u_R} =& - \frac{g_s^2\, \mathcal{C}(\mathbf{3})  + g'^2 \, y_{u_R}^2}{16\pi^2 \varepsilon}C_{u_R} - \frac{Y_u^\dag  C_{q_L} Y_u}{16\pi^2 \varepsilon} + \frac{Y_u^\dag Y_u }{16\pi^2 \varepsilon} C_H \, , \\
	\delta C_{d_R} =& - \frac{g_s^2\, \mathcal{C}(\mathbf{3})  + g'^2 \, y_{d_R}^2}{16\pi^2 \varepsilon}C_{d_R} - \frac{Y_d^\dag  C_{q_L} Y_d}{16\pi^2 \varepsilon} + \frac{Y_d^\dag Y_d }{16\pi^2 \varepsilon} C_H \, ,\\
	\delta C_{\ell_L} =& - \frac{g^2\, \mathcal{C}(\mathbf{2})  + g'^2 \, y_{\ell_L}^2}{16\pi^2 \varepsilon}C_{\ell_L} - \frac{Y_e^\dag  C_{e_R} Y_e}{32\pi^2 \varepsilon} + \frac{Y_e^\dag Y_e }{32\pi^2 \varepsilon} C_H \, ,\\
	\delta C_{e_R} =& - \frac{g'^2 \, y_{e_R}^2}{16\pi^2 \varepsilon}C_{e_R} - \frac{Y_e^\dag  C_{\ell_L} Y_e}{16\pi^2 \varepsilon} + \frac{Y_e^\dag Y_e }{16\pi^2 \varepsilon} C_H \, , \\
	\delta C_H =& +\frac{g^2\, \mathcal{C}(\mathbf{2}) + g'^2\, y_H^2}{8\pi^2\varepsilon} + \frac{3 \mathrm{tr}(C_{q_L} \hat{Y}_q^2)}{16\pi^2 \varepsilon}\\
	& \qquad {} -\frac{3\left(\mathrm{tr}(Y_u C_{u_R} Y_u^\dag) - \mathrm{tr}(Y_d C_{d_R} Y_d^\dag) \right)}{16\pi^2 \varepsilon} - \frac{ \mathrm{tr}(Y_e^\dag C_{\ell_L} Y_e) - \mathrm{tr}(Y_e C_{e_R} Y_e^\dag)}{16\pi^2 \varepsilon} \, .
\end{aligned}\end{align}
Notice that, in addition to the counterterms of Eq.~\eqref{eq:ct_vertex_above_mZ}, also the redundant operator $\partial_\nu B^{\nu\mu}$ is generated via the loops of Fig.~\ref{fig:redop_corr}, and a further counterterm $\delta C_B$ is needed. We get
\begin{align}\begin{aligned}
	\delta C_B &= - \frac{2}{3} \frac{g'}{16\pi^2 \varepsilon} T\, ,
\end{aligned}\end{align}
with $T$ defined in Eq.~\eqref{eq:T}. Once $\delta C_B$ is added to the Lagrangian, we apply Eq.~\eqref{eq:EOM_1} to define away the $\partial^\nu B_{\nu\mu}$ current, obtaining that each of the counterterms in Eq.~\eqref{eq:ct_vertex_above_mZ} gets a correction
\begin{align}
	\delta C_a \to \delta C_a - y_a \frac{2}{3} \frac{g'^2}{16\pi^2 \varepsilon} T\, .
\end{align}
We are now in the position of finally compute the RGE's of the Wilson coefficients $C_\varphi$. Let us sketch the procedure. Writing $\varphi \simeq (\mathbb{1} + W_\varphi) \varphi_{bare}$ for each field (with explicit expressions given in Eq.~\eqref{eq:wavefunction}), the bare Wilson coefficient is given in terms of the renormalized one by
\begin{align}\label{eq:bare_ren_coupling}
	C_\varphi^{bare} = \mu^{\alpha \varepsilon}(\mathbb{1} + W_\varphi)(C_\varphi + \delta C_\varphi)(\mathbb{1} + W_\varphi) \simeq \mu^{\alpha \varepsilon} \left( C_\varphi +\delta C_\varphi + \frac{C_\varphi W_\varphi + W_\varphi C_\varphi}{2}\right) \, .
\end{align}
The factor $ \mu^{\alpha \varepsilon}$ is inserted to ensure that all the renormalized Wilson coefficients $C_\varphi$ are dimensionless in $d= 4-2\varepsilon$ dimensions. The coefficient $\alpha$ depends on the field content of the dark current $J_{\cal D}$, but we will not need to specify it as long as all the SM currents couple either to the same dark current, or to many dark currents of the same dimensions. Imposing $dC_\varphi^{bare}/d\log\mu = 0$ and using that, to leading order in the couplings, the RGE's have the form
\begin{align}\begin{aligned}\label{eq:useful_app}
	\frac{d C_\varphi}{d\log\mu} = -\alpha\, \varepsilon\, C_\varphi + \dots \, , 
\end{aligned}\end{align}
we obtain the RGE's of Eq.~\eqref{eq:RGE_above}. For the running of the Yukawa and gauge couplings we use the results in References~\cite{Machacek:1983fi,Machacek:1983tz}. \vv

Let us now move to the EFT \textit{below} the EW scale. The only contributions to the wave function renormalization of fermions come from QED and QCD. The counterterms are
\begin{align}\begin{aligned}
	\mathcal{L}_{c.t.} =  - \frac{1}{16\pi^2 \varepsilon} \sum_f \left( \mathcal{C}_f(\mathbf{3}) g_s^2 + Q_f^2 e^2 \right) \overline{f}  i \slashed{\partial} f\, ,
\end{aligned}\end{align}
where $\mathcal{C}_f(\mathbf{3})$ is the $SU(3)_c$ quadratic Casimir, if the fermion has color, and $Q_f$ is the fermion electric charge. Notice that there is no flavor off-diagonal contribution. Turning to vertex corrections, we now have contribution from gauge bosons and from four fermion interactions (see Fig.~\ref{fig:vertex_below}). The vertex counterterms are
\begin{align}\begin{aligned}
\delta C_{V_u}  & = \frac{1}{16\pi^2\epsilon}\left[\big(\mathcal{C}_2(3) g_s^2 + Q_u^2 e^2\big) C_{V_u} - \frac{g_{V_u}\mathcal{F}_u}{2} - \frac{{G_F}_{ud}}{2}V \left(\mathcal{M}_{V_d}^2- \mathcal{M}_{A_d}^2 \right) V^\dagger\right]\, ,
\\
\delta C_{V_d}  & = \frac{1}{16\pi^2\epsilon}\left[\big(\mathcal{C}_2(3) g_s^2 + Q_d^2 e^2\big) C_{V_d} -\frac{g_{V_d}\mathcal{F}_d}{2} - \frac{{G_F}_{du}}{2}V^\dagger \left(\mathcal{M}_{V_u}^2 - \mathcal{M}_{A_u}^2 \right)V \right]\, ,
\\
\delta C_{V_\nu}  & = \frac{1}{16\pi^2\epsilon} \left[ -\frac{g_{V_\nu}\mathcal{F}_\nu}{2} - \frac{{G_F}_{\nu e}}{2}(\mathcal{M}_{V_e}^2-\mathcal{M}_{A_e}^2)\right]\, ,
\\
\delta C_{V_e}  & = \frac{1}{16 \pi^2 \epsilon} \left[ Q_e^2 e^2 C_{V_e}-\frac{g_{V_e}\mathcal{F}_e}{2} \right]\, ,
\\
\delta C_{A_u}  & = \frac{1}{16\pi^2\epsilon}\left[\big(\mathcal{C}_2(3) g_s^2 + Q_u^2 e^2 \big) C_{A_u} -\frac{g_{A_u}\mathcal{F}_u}{2} + \frac{{G_F}_{ud}}{2}V \left(\mathcal{M}_{V_d}^2 -\mathcal{M}_{A_d}^2 \right)V^\dagger\right]\, ,
\\
\delta C_{A_d}  & =  \frac{1}{16\pi^2\epsilon}\left[\big(\mathcal{C}_2(3) g_s^2 + Q_d^2 e^2 \big) C_{A_d} - \frac{g_{A_d}\mathcal{F}_d}{2} + \frac{{G_F}_{du}}{2}V^\dagger \left(\mathcal{M}_{V_u}^2 - \mathcal{M}_{A_u}^2 \right) V\right]\, ,
\\
\delta C_{A_\nu}  & =  \frac{1}{16\pi^2\epsilon} \left[ -\frac{g_{A_\nu}\mathcal{F}_\nu}{2} + \frac{{G_F}_{\nu e}}{2}(\mathcal{M}_{V_e}^2-\mathcal{M}_{A_e}^2)\right]\, ,
\\
\delta C_{A_e}  & = \frac{1}{16\pi^2 \epsilon} \left[ Q_e^2 e^2 C_{A_e}-\frac{g_{A_e}\mathcal{F}_e}{2}\right]\, ,
\label{Eq:ct_below_EW}
\end{aligned}\end{align}
where we have used the definition of Eq.~\eqref{eq:useful_below}. As before, the redundant current $\partial_\nu F^{\nu\mu}$ is radiatively generated, and the corresponding counterterm results
\begin{align}
	\delta C_\gamma = \frac{e}{12 \pi^2 \varepsilon} \big(3 Q_u \mathrm{tr}[C_{V_u}] + 3 Q_d \mathrm{tr}[C_{V_d}] + Q_e \mathrm{tr}[C_{V_e}]  \big)\, .
\end{align}
This effect can be incorporated in the other Wilson coefficients via the shift
\begin{align}
	C_{V_a} \to C_{V_a} + Q_a \mathcal{Q}\, .
\end{align}
Repeating now the procedure sketched in Eqs.~\eqref{eq:bare_ren_coupling} and~\eqref{eq:useful_app} we obtain the RGE's presented in Eq.~\eqref{eq:RGE_below}.

\section{Running of $G_F$ due to QCD and QED}\label{app:running_Gf}
\begin{figure}[t!b]
	\centering
	\includegraphics[width = .2 \textwidth]{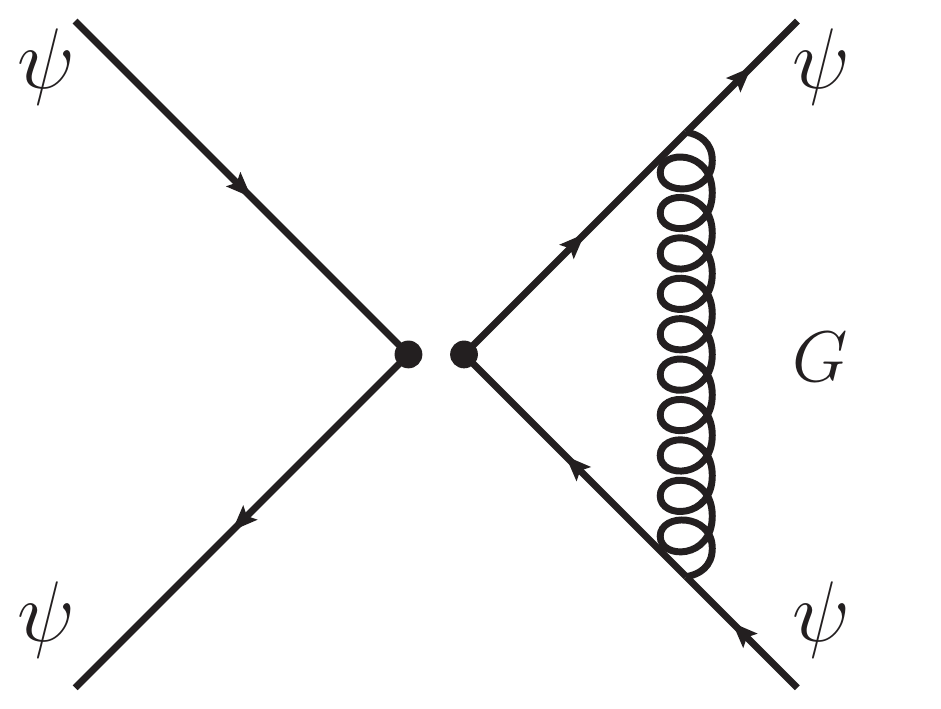} 
	\includegraphics[width = .2 \textwidth]{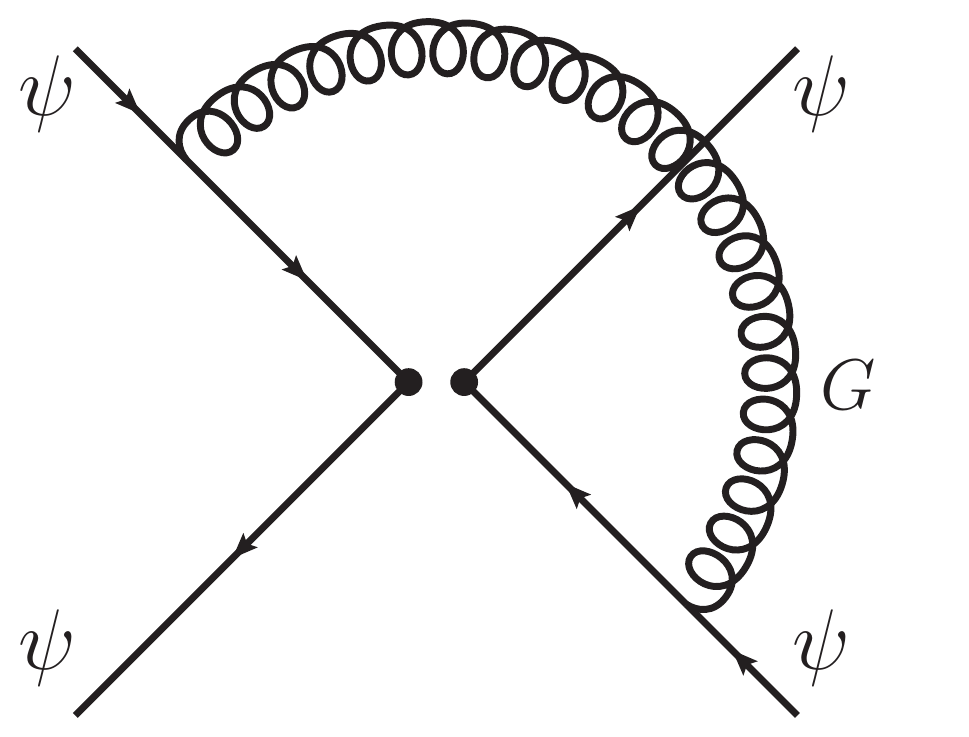}
	\includegraphics[width = .2 \textwidth]{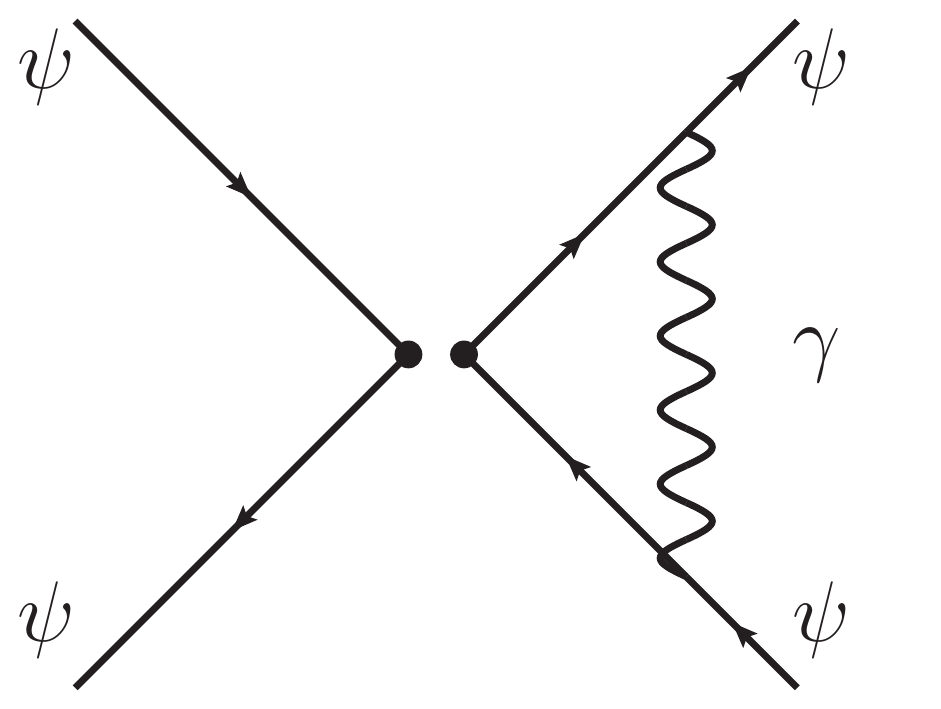}
	\includegraphics[width = .2 \textwidth]{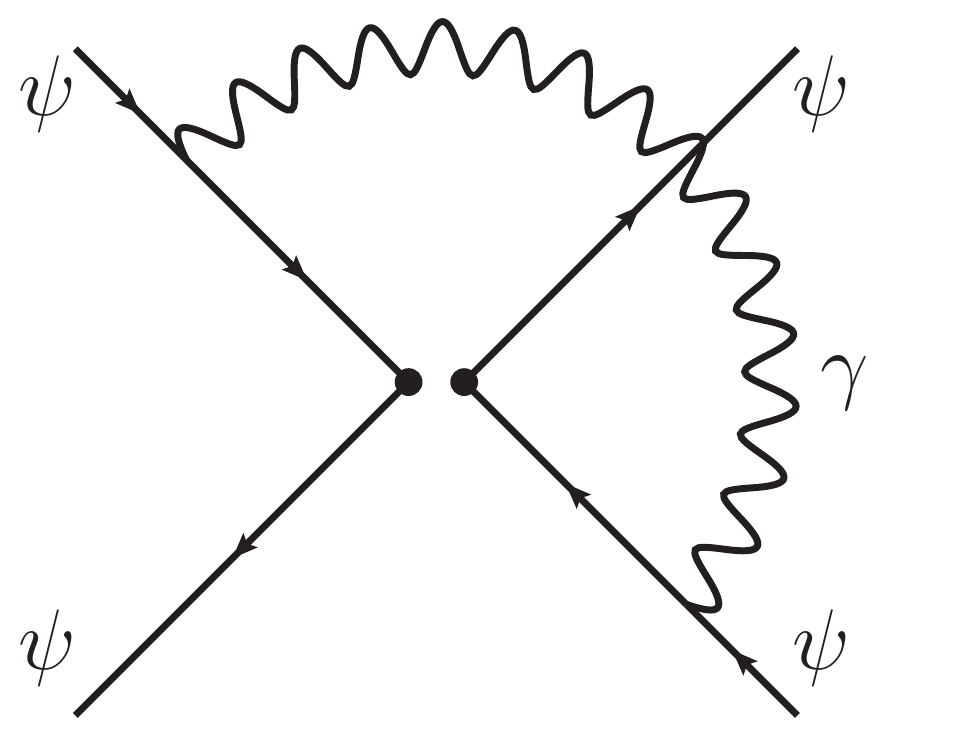}
	\caption{\label{fig:GFermi} Feynman diagrams contributing to the running of the Fermi coupling below the EW scale.}
\end{figure}
Let us now discuss the running of the Fermi coupling below the EW scale. In addition to the wave function renormalization, we need to consider the diagrams shown in Fig.~\ref{fig:GFermi}. The results are independent on the flavor of the external fermions. Some of the vertex corrections will cancel against the wave function contributions (more specifically, those vertex corrections in which the gluon or photon connect particles in the same fermion line). The gluon `crossed' contributions~\footnote{With `crossed' contributions we mean those loops in which the photon or gluon connect particles in two different fermion lines.} generate the operators
\begin{align}
	\overline{q} \Gamma^\mu T^a q \overline{q'} \Gamma'_\mu T^a q'\, 
\end{align}
with a double insertion of Gell Mann matrices. These operators do not enter in the running of our currents, and we will therefore not consider them in the following. Notice however that their effect is important when the nature of the dark current is specified and dark fermions loops can be considered, as in Refs.~\cite{Brod:2018ust,Bishara:2018vix}. We are thus left with the photon `crossed' loops, which are the only radiative effects that we need to take into account. Their effect is to produce the RGE
\begin{align}\label{eq:Grunning}
\frac{d {G_F}_{ff'}}{d\log\mu} = -2 {G_F}_{ff'} \frac{Q_f Q_{f'}\alpha}{ \pi },
\end{align}
which has been used in the numerical computations of Sec.~\ref{sec:results}. Notice that the resulting RGE is independent on $\Gamma^\mu$ and $\Gamma'^\mu$. Quantitatively, the relative variation in the value of ${G_F}_{ff'}$ is of order of a few percent.

\bibliography{flavor_dark_sector}{}
\bibliographystyle{JHEP}

\end{document}